\documentclass[useAMS,usenatbib]{mn2e}

\usepackage{longtable}
\usepackage{graphicx}

\newcommand{\msun}{\mbox {$\rm M_{\odot} $}}

\usepackage{color}

\title[Low Mass Stars and Brown Dwarfs In Praesepe]{Low Mass Stars and Brown Dwarfs in Praesepe} 
\author[D.E.A. Baker]{D.E.A. Baker$^{1}$\thanks{E-mail:
deab1@star.le.ac.uk}, R.F. Jameson$^{1}$, S.L. Casewell$^{1}$, N. Deacon$^{2,3}$, N.Lodieu$^{4,5}$ and \and N. Hambly$^{6}$\\
$^{[1]}$Department of Physics and Astronomy, University of Leicester, University Road, Leicester, LE1 7RH, U.K.\\
$^{[2]}$Department of Astrophysics, Faculty of Science, Radboud University Nijmegen, PO Box 9010, 6500 GL Nijmegen, The Netherlands\\
$^{[3]}$Institute for Astronomy, University of Hawaii, 2680 Woodlawn Drive, 96822, Honolulu, Hawaii, USA\\
$^{[4]}$Instituto de Astrof\'isica de Canarias, C/ v\'ia L\'acta s/n, E-38200 La Laguna, Tenerife, Spain\\
$^{[5]}$Departamento de Astrof\'isica, Universidad de La Laguna, E-38250 La Laguna, Tenerife, Spain\\
$^{[6]}$Institute for Astronomy, SUPA (Scottish Universities Physics Alliance), University of Edinburgh, Royal Observatory,\\   Blackford Hill, Edinburgh, EH9 3HJ, U.K.}
\begin{document}

\date{30$^{th}$ June 2010}

\pagerange{\pageref{firstpage}--\pageref{lastpage}}
\pubyear{2010}

\maketitle

\label{firstpage}

\begin{abstract}
Presented are the results of a large and deep optical-near-infrared multi-epoch survey of the Praesepe open star cluster using data from the UKIDSS 
Galactic Clusters Survey. Multiple colour magnitude diagrams were used to select potential members and proper motions were used to assign levels of 
membership probability. From our sample, 145 objects were designated as high probability members ($p \geq 0.6$) with most of these having been found by 
previous surveys although 14 new cluster members are also identified. Our membership assignment is restricted to the bright sample of objects ($Z < 18$). 
From the fainter sample, 39 candidates were found from an examination of multiple colour magnitude plots. Of these, 2 have small but significant 
membership probabilities. Finally, using theoretical models, cluster luminosity and mass functions were plotted with the later being fitted with 
a power law of $\alpha=1.11 \pm 0.37$ for the mass range 0.6 to 0.125 \msun and an assumed cluster age of 500 Myrs in the UKIDSS $Z$ photometric band. 
Likewise taking an assumed cluster age of 1 Gyr we find $\alpha=1.10 \pm 0.37$. Similar values were also found for the $J$ and $K$ bands. 
These results compare favourably with the result of \cite{Kraus_2007} ($\alpha=1.4 \pm 0.2$) but are significantly lower than that of the more recent study conducted 
by \cite{Boudreault_2009} ($\alpha=1.8 \pm 0.1$). 
\end{abstract}


\begin{figure*}
\begin{center}
\includegraphics[width=\columnwidth]{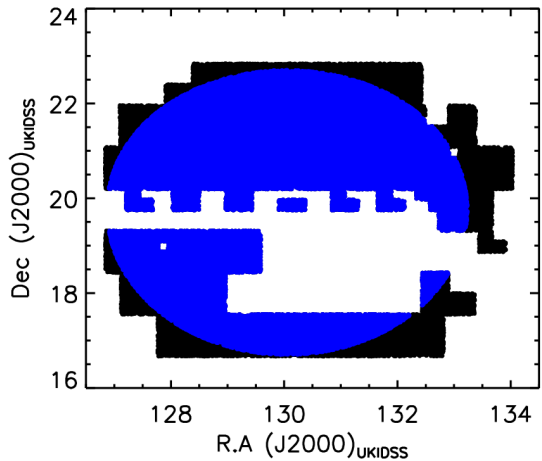}
\caption{The full coverage of the Praesepe star cluster available from UKIDSS DR6 with the blue 
region denoting the sources present in the 3 degree radial selection.}
\label{survey_area}
\end{center}
\end{figure*}

\section{Introduction}
Low mass stars (LMS) and Brown Dwarfs (BDs) are the lowest mass objects for which the stellar formation process is applicable (\citealt{Burrows_2001}). 
Locating these objects in a cluster is particularly important as it provides some of the key parameters that help define them, notably age, distance 
and metallicity. These can in turn be used to help provide constraints to theoretical evolutionary and atmospheric models. Another important 
reason for trying to locate these objects in clusters, particularly older ones, is that as the cluster ages it undergoes a process of dynamical 
evolution, in which the lower mass objects are preferentially ejected from the cluster into the field (\citealt{de_la_fuente_marcos_2000}). 
This ejection of objects causes changes to occur in the cluster's luminosity and mass function, which for a gravitationally bound association 
can be considered a proxy for the initial mass function (IMF) of the system. Although there have been many previous studies aimed at characterising 
the IMF, they have often come from many surveys conducted in different filters and with different instruments. For this reason the UKIDSS Galactic 
Clusters Survey (GCS; \citealt{Lawrence_2007}) was devised. One of the clusters surveyed is the open star cluster of Praesepe.

Praesepe lies at a distance of $\approx$180 pc (($M-m$)$_{0}=$6.30 $\pm$ 0.003, \citealt{van_leeuwen_2009}) with zero 
reddening and near solar metallicity. Whilst the distance is fairly well constrained, there is a lack of agreement with 
regards to its age. \cite{Allen_1973} placed the value of Praesepe's age towards the lower end of the scale at 430 Myr 
and it was the standard value for many years. Later work by \cite{Vandenberg_1984} placed it at a much higher value of 900 Myr. 
Their value was obtained via the fitting of models describing the main-sequence to the observed colour magnitude diagrams (CMDs). 
The widest range of age estimates is reported by \cite{Tsvetkov_1993} who placed a similar lower limit of 540 Myr but an 
upper limit of over 1.5 Gyrs. The method of \citeauthor{Tsvetkov_1993} does not depend on the fitting of the Zero-Age-Main-Sequence 
but instead relied on models used to calculate the ages of $\delta$ Scuti stars that are present within 
the cluster. Finally, \cite{Kharchenko_2005} placed the age in the middle at 795 Myrs after using the Padova grid of post main sequence 
isochrones from \cite{Girardi_2002}.
We consider this to be closest to the most likely age as there is some evidence that the Hyades cluster at 625 Myrs shares a common origin with 
Praesepe (\citealt{Eggen_1960}; \citealt{Henry_1977}). However due to the uncertainty that exists, two ages for the cluster have been 
adopted throughout this paper. The first is an age of 500 Myr and the second 1 Gyr. These were chosen as they coincide nicely with ages for 
which the NextGen (\citealt{Baraffe_1998}) and DUSTY (\citealt{Chabrier_2000}) theoretical models have been calculated. In this work we have 
used the BT-NextGen and BT-DUSTY models which are based on the aforementioned versions 
except they have been calculated with updated opacity data\footnote{http://phoenix.ens-lyon.fr/simulator/index.faces}.

Praesepe's members all have a common proper motion centred around $\mu_\alpha=-$35.81 mas yr$^{-1}$ and $\mu_\delta=-$12.85 mas yr$^{-1}$, again from 
the work by \citeauthor{van_leeuwen_2009} based on re-reduced Hipparcos data. This distinct proper motion allows relatively 
easy photometric and astrometric membership surveys to take place. The ``high-mass'' stellar 
population ($V$ $<$ 13) was identified by \cite{Klein-Wassink_1927} with ``intermediate-mass'' ($V$ $<$ 17) and ``low-mass'' 
M-dwarfs ($R$ $>$ 20) being identified by \cite{Jones_1983} and \cite{Hambly_1995a} respectively. Further work has been carried 
out by \citeauthor{Pinfield_1997} (1997, 2003), \cite{Adams_2002}, \cite{Chappelle_2005} and \cite{Gonzalez_2006}. However, 
these surveys have often proved to be contaminated with an excess of field stars, as in the case of \citeauthor{Adams_2002} 
or have no proper motion information (\citealt{Pinfield_1997}). The most comprehensive study to date was produce by \cite{Kraus_2007}, who 
used data from the Sloan Digital Sky Survey (SDSS; \citealt{York_2000}), Two Micron All Sky Survey (2MASS; \citealt{Skrutskie_2006}), 
USNOB1.0 (\citealt{Monet_2003}) and finally UCAC2 (\citealt{Zacharias_2004}) to find 1010 candidate members of Praesepe, 
442 being identified for the first time, down to a spectral type of around M5. The work presented in this paper is 
again a return to a search for this ``low-mass'' population, however we have based our search on data made available from 
the UKIRT Infrared Deep Sky Survey (UKIDSS), 2MASS and SDSS.

\section{The Surveys}
UKIDSS is a near-infrared sky survey that aims to survey some 7500 square degrees of the northern
sky within its science operation lifetime. The depth that it aims to achieve will be three magnitudes deeper 
than that offered by 2MASS, making UKIDSS an ideal companion to the SDSS in areas where the two coincide. 
UKIDSS uses the United Kingdom Infrared 3.8m Telescope (UKIRT) located on Mauna Kea and the Wide Field CAMera (WFCAM). 
WFCAM itself consists of four Rockwell Hawaii-II
(HgCdTe) detectors, each of dimension 2048 x 2048 pixels. A single
pixel represents a scale of 0.4 arc seconds and the spacing between the
detectors requires that four paw-prints be undertaken in order to
construct a single 0.8deg$^{2}$ tile.
The UKIDSS survey program consists of five separate components. The Large
Area Survey, The Galactic Plane Survey, The Deep
Extragalactic Survey, The Ultra Deep Survey and The
Galactic Clusters Survey, which provides the focus and data for
this paper.For more specific information on the UKIDSS programme a set
of reference papers have been produced which aim to provide
technical documentation on the infrared survey instrument itself
(WFCAM; \citealt{Casali_2007}), the WFCAM photometric system 
(\citealt{Hewett_2006}; \citealt{Hodgkin_2009}), the UKIDSS surveys 
(\citealt{Lawrence_2007}), the pipeline processing system 
(\citealt{Irwin_2008}) and finally the science archive as described in \cite{Hambly_2008}.

The GCS aims to enable a comprehensive study of 10
star-forming regions and clusters, with hopes of detailing the form of the IMF 
and how it is affected by the environment in the sub-stellar regime. 
All the data for these programmes is processed
by the Cambridge Astronomical Survey Unit (CASU) and then archived and
released by the WFCAM Science Archive (WSA) located in
Edinburgh. Currently, UKIDSS is on data release 6 (DR6) as of $13^{th}$ October 2009
for the ESO community (and DR3 as of the
$5^{th}$ June 2009 for the world release). DR6  has reported depths of $Z=20.4$,
$Y=20.1$, $J=19.6$, $H=18.8$ and $K=18.2$ (first epoch). At the time of
writing the Praesepe cluster has not yet been fully surveyed with only
$\approx$ 23 square degrees being available in all filters and imposing a 3
degree radial selection from the cluster centre $\approx$ 18 square
degrees. This can be clearly seen in Figure~\ref{survey_area}. Due to the missing
region consisting of mainly the cluster centre, a clear lack of overlap
will exist between this and any of the previous bodies of work. As such,
this work fails to retrieve many of the previously identified cluster
members and so acts to serve as an incremental part of a full cluster survey 
within the UKIDSS programme.

To retrieve the data from the WSA a similar SQL query to that 
of \citet{Lodieu_2007a} was devised (See Appendix~\ref{SQL_query} for the full queries). The query was
adapted to cross match with the Sloan Sky Survey through the use
of the newly implemented gcsSourceXDR7PhotoObj table. The Class
parameter in each of the five filter bands was set to only select
objects that matched with criteria -2 or -1 in value, i.e. those that
had been deemed stellar in nature by the pipeline. While this
clearly limits the number of sources by requiring the object to be
present in all bands, particularly at the faint end \citep{Lodieu_2007b}, 
it does mean a greater level of reliability for the data that have been
selected. Alongside the class selection criteria, the query also placed various 
quality control mechanisms as defined by the use of the post processing 
error bits flags on the UKIDSS data and the flags contained within the SDSS
subsection\footnote{The SDSS flag selections were taken from clean photometry 
section of the SDSS SQL query sample page 
http://cas.sdss.org/dr6/en/help/docs/realquery.asp Due to the nature of the 
objects being investigated the constraints were placed only on bands in which the 
object was likely to be present i.e The i and z$'$ bands.}.   

The SQL query retrieved a total of 79,162 sources from the
archive. When asking for the 2MASS and SDSS cross tables, the UKIDSS
source identifier was used in order to merge the two separate queries 
into a master table, with each source containing any 
data from UKIDSS, 2MASS and/or SDSS. This match was performed
using the TOPCAT program in the STARLINK suite of programs. To try and minimise the 
contamination due to field stars at the outer edges of the survey area where the cluster is more 
diffuse we employed a radial cut of 3 degrees from the cluster centre. This left 59,779 sources, which we call our GCS dataset.

\begin{figure}
\begin{center}
\includegraphics[width=\columnwidth]{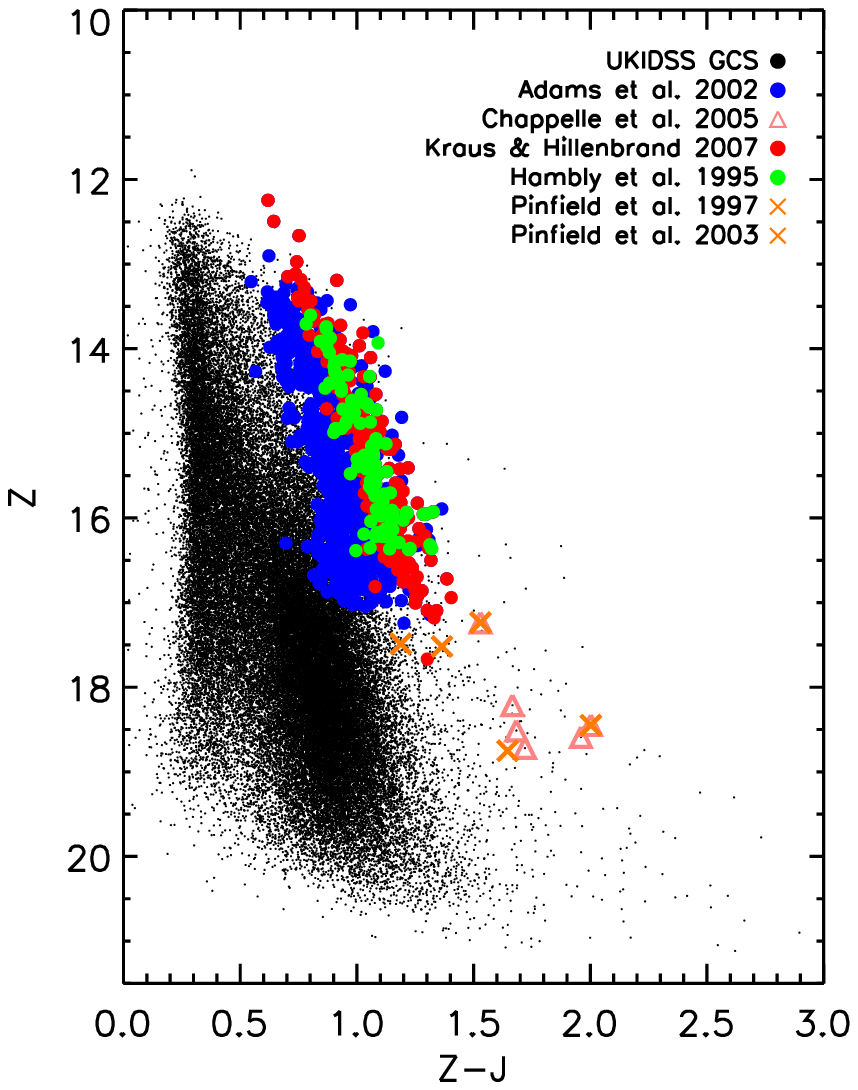}\label{GCS_plus_matches_cmd}
\caption{($Z$-$J$,$Z$) CMD for the $\approx$ 18 square degrees of the
Praesepe cluster selected from the WFCAM Science Archive with a
corresponding 2MASS source within the inner 3 degrees of the cluster
centre. Matches to this data set from the surveys of \citeauthor{Adams_2002} (\citeyear{Adams_2002}), \citeauthor{Chappelle_2005} (\citeyear{Chappelle_2005}),
\citeauthor{Hambly_1995b} (\citeyear{Hambly_1995b}), \citeauthor{Kraus_2007} (\citeyear{Kraus_2007}), and \citeauthor{Pinfield_2003} (\citeyear{Pinfield_1997} and \citeyear{Pinfield_2003}) are also shown. The survey of \citeauthor{Gonzalez_2006} 
did overlap with this area however for those sources within the region no
matches were found, due possibly to our strict selection criteria on
the UKIDSS data.}
\end{center}
\end{figure}

The objects found by
\cite{Adams_2002}, \cite{Chappelle_2005}, \cite{Hambly_1995b}, \cite{Gonzalez_2006}, \cite{Kraus_2007}, \cite{Pinfield_1997} and \cite{Pinfield_2003}
were then matched to this GCS dataset to select only those whose
survey areas overlapped and could be recovered from our data. In total
642 sources were recovered from \citeauthor{Adams_2002} (who in total report 4,954 objects for their whole survey), 6 from \citeauthor{Chappelle_2005} (26 in total), 109 from
\citeauthor{Hambly_1995b} (515 in total), 0 from \citeauthor{Gonzalez_2006} (20 in total), 274 from \citeauthor{Kraus_2007} (1,130 in total), and 5 from each of the
Pinfield surveys\footnote{The objects retrieved from the two Pinfield surveys are the same 5 objects.} 
these can all be seen in Figure~\ref{GCS_plus_matches_cmd}.

The first run of the SQL query contained a
cross-correlation between the UKIDSS DR6 GCS dataset with its nearest
2MASS counterpart. This cross-correlation allows a determination of proper motion for the
matched objects. Typically over a small area the astrometry provided
by 2MASS is good to 50 mas \citep{Skrutskie_2006}. The CASU
pipeline performs its astrometric calibration for the WFCAM data based
on point sources within the 2MASS catalogues. Hence, accurate relative proper
motions can be derived by simply taking the difference in 2MASS and
WFCAM positions and dividing by the epoch difference (\citealt{Lodieu_2007a}; \citealt{Jameson_2008a}). This was
automatically done and the results converted \hbox{into mas yr$^{-1}$} by the SQL
query when run through the WSA data centre. The proper motions are described as relative as
they exhibit a distinct movement in contrast to the comparably stationary background.
The accuracy of the astrometry and the average time baseline of around 5 years provides an error \hbox{of $\approx$10 mas
  yr$^{-1}$}. Of the 59,779 sources 34,990 were found to have 2MASS counterparts leaving 24,789 with no 2MASS identifier. 
Because 2MASS lacks the depth of UKIDSS we only retrieved the brighter of our sample ($K < 16.5$) from this dataset, with the SDSS
dataset providing the fainter candidates (The extent of the data set can be seen in Figure~\ref{GCS_plus_matches_cmd}).

\begin{figure*}
\begin{center}
\includegraphics[width=2\columnwidth]{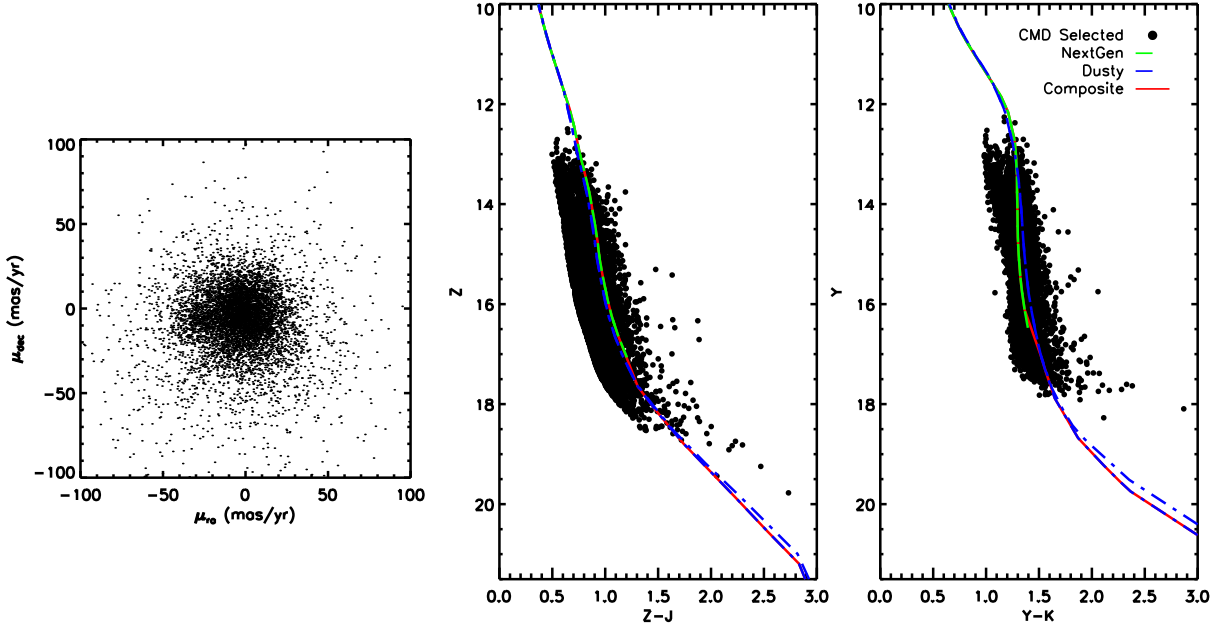}\label{VPD_CMD_Selected}
\caption{On the left is the VPD for the sources
selected from our spatial and colour cuts in the ($Z$-$J$,$Z$) and ($Y$-$K$,$Y$)
colour magnitude diagrams as shown in the middle and on the right
respectively. The cluster can be seen as the over-density of objects around -30 and -10 mas
yr$^{-1}$ in right ascension and declination. The green dashed line is
the theoretical isochrone for the BT-NextGen evolutionary model and the
blue the BT-DUSTY equivalent. Both the 500 Myr and 1 Gyr flavours have been
considered due to the uncertainty in the age of Praesepe.}
\end{center}
\end{figure*}

Thanks to the newly implemented gcsSourceXDR7PhotoObj table linking
the WFCAM DR6 and SDSS DR7 data sets at the WSA,
a cross-correlation between the UKIDSS data set and that of the SDSS
was also available for interrogation. SDSS DR7 reported having
\hbox{surveyed 11,000 square degrees} in all of its five filters
($\textit{ugriz$'$}$) which included the full area of Praesepe
\citep{Kraus_2007}. Upon inspection of the survey dates it became
apparent that only a short amount of time had elapsed between the survey of
Praesepe by Sloan and that of UKIDSS ($\approx$ 2 - 2.5 years on average). The lack of a decent baseline therefore warranted a different approach in order to calculate 
proper motions and thus cluster membership assignment from that which will be detailed for 2MASS. 
Again the 59,779 UKIDSS sources were retrieved with 53,562 having an SDSS counterpart and 6,253 being
unique to UKIDSS. In total UKIDSS has 2,225 unique sources for which no
counterpart has been found in either 2MASS of SDSS (30,998 objects were present in all three surveys).

\section{Members of Praesepe from 2MASS}

This section will describe the processes undertaken to construct the
list of candidate cluster members in Praesepe from the UKIDSS colour
magnitude diagrams, and where possible proper motion vector point
analysis based from the 2MASS-UKIDSS cross-correlation. The
procedure is as follows:

\begin{enumerate} 
\item Select only those sources that have a 2MASS identifier
associated with their UKIDSS identifier and are within 3 degrees of
the defined cluster centre (34,990 objects).
\item To check the proper motion errors fit the reduced data set (CMD and radius selections imposed)
with a two dimensional Gaussian, then use the $\sigma$ of the
Gaussian to act as a proxy for the error.
\item Using the theoretical isochrones, select objects 
that are no more than 0.3 magnitudes to the left, and all of those on 
the right in both the ($Z$-$J$,$Z$) and ($Y$-$K$,$Y$) CMDs.
\item Analyse the resulting Vector Point Diagram (VPD) for this colour
selected data set across a range of magnitude bins and inferring from the
probability fitting routine a level of cluster membership for each
object. Then selecting out a high probability sample (HPM) as those objects with
assigned probabilities p $\geq$ 0.60.
\item Again look at this HPM sample and reject any obvious
non-photometric candidates.
\end{enumerate}

\subsection{Calculating Proper Motions} 
The cross-correlation of UKIDSS with 2MASS provided a value for the 
matched objects proper motion in $\mu_\alpha\cos_\delta$ and
$\mu_\delta$. An estimate of the errors based on the time baseline
and specifications of 2MASS is $\approx$ 10 mas
yr$^{-1}$. To confirm this error estimate, the proper motions for those
sources that lay within 3 degrees of the cluster centre (which
were spread over a region of proper motion space from -150 to 150 mas
yr$^{-1}$) were divided into bins of 20 mas yr$^{-1}$ and the number
in each bin totalled. A two dimensional Gaussian was then fitted enabling, 
a determination of the cluster spread. Objects that were
defined as being outside of the 3$\sigma$ limit were then removed and
the fit reapplied. The $\sigma$ of the Gaussian then provides an
estimate for the error in the proper motions \citep{Jameson_2008a} and
was found to be of the order of $\approx$ 12 mas yr$^{-1}$, instead of our assumed 10 mas yr$^{-1}$.

\subsection{Colour Magnitude Diagrams}  
In order to select candidates from the CMDs we made
use of the BT-NextGen and BT-DUSTY theoretical isochrones which have had the distance modulus of the
cluster added to them in order. The models represent different evolutionary paths
for describing objects with and without ``dusty'' atmospheres. Due to 
the temperatures and masses being explored by this survey the
objects involved inhabit both of these regimes and so the two
isochrones need to be combined. To create a composite isochrone line,
the data points from the BT-NextGen isochrone in the relevant band filter
were taken to their minimum temperature $\approx$ 3000 K. This temperature lies just above where dust grains begin to form in the BD 
atmosphere and the BT-NextGen models become invalid. We thus combine the BT-DUSTY isochrone models ($T_{eff} < 3000$ K)  with the 
BT-Nextgen models at this point with a simple straight line added in between the resulting break. This composite
isochrone was calculated for both the 500 Myr and 1 Gyr evolutionary
models due to the uncertainty in the age of Praesepe. It was then employed in both 
the ($Z$-$J$,$Z$) and ($Y$-$K$,$Y$) CMDs with sources laying no more than 0.3 magnitudes to the left in the
horizontal direction from the line and those to the right being
selected and passed to the proper motion fitting routine. See Figure~\ref{VPD_CMD_Selected}
for the VPD and the two CMDs
associated with the spatial and colour cut selections. This process
selected 7,127 objects out of the possible 34,990. This selection is
rather conservative as it aimed to include all the cluster members
from the previous surveys, most notably that of \cite{Adams_2002} whose
objects appear far bluer than those found by  both \cite{Kraus_2007} and
by \cite{Hambly_1995b}.

\section{Cluster Distributions and Membership Probability}

To calculate the membership probabilities for the data sample two
distributions were fitted to the  cluster. One, a circularly symmetric
Gaussian as originally employed by \citet{Sanders_1971} and also
\citet{Francic_1989}, and two, an exponential decaying in the
direction of the cluster proper motion centre coupled with a
perpendicularly oriented Gaussian (\citealt{Hambly_1995b};
\citealt{Deacon_2004}). For the fitting process to work the cluster
proper motion centre  (-35.81,-12.85) is rotated from its original
position on the Vector Point Diagram to lie on the  y-axis. The
following set of equations describe the field distribution ($\Phi_{f}$), cluster
distribution ($\Phi_{c}$) and the normalisation to the exponential ($c_{o}$).

\begin{equation} \Phi_{f} =
	\frac{c_{o}}{\sqrt{2\pi}\Sigma_{x}}exp\left(-\frac{(\mu_{x}-\mu_{xf})^{2}}{2\Sigma_{x}^{2}}-\frac{\mu_{y}}{\tau}\right).
	\end{equation}

\begin{equation} \Phi_{c} =
	\frac{1}{2\pi\sigma^{2}}exp\left(-\frac{(\mu_{x}-\mu_{xc})^{2}+(\mu_{y}-\mu_{yc})^{2}}{2\sigma^{2}}\right).
	\end{equation}

\begin{equation}
c_{o}=\frac{1}{\tau\left(e^{-\frac{\mu_{1}}{\tau}}-e^{-\frac{\mu_{2}}{\tau}}\right)}.
\end{equation}
Where,

\begin{equation}
c_{o}\int\limits_{\mu_{1}}^{\mu_{2}}e^{-\frac{\mu_{y}}{\tau}}d\mu_{y}=1.
\end{equation} 

The values of $\mu_x$ and $\mu_y$ refer to the proper motion attributed to each individual object. The quantity $\sigma$ is the Gaussian width, 
whilst $\mu_{xc}$ and $\mu_{yc}$ are the cluster's mean proper motion. $\Sigma_{x}$ is the proper motion dispersion value in x and $\tau$ the 
exponential scale length for the field proper motion distribution in $y$. $\mu_{xf}$ is the field mean proper motion in $x$ and finally $\mu_{1}$ 
and $\mu_{2}$ are the limits for the normalisation to the exponential $c_{o}$. For the rotated VPD these were set at 20 and 70 mas yr$^{-1}$ to 
avoid the mass of stars centred around (0,0). 

Combining the field star distribution and the cluster distribution 
with information about the fraction of stars which are field
stars ($f$) the resulting expression for the total distribution ($\Phi$) is

\begin{equation} \Phi=f\Phi_{f}+(1-f)\Phi_{c}. \end{equation}

After employing the method of maximum likelihood with $\Theta$ representing one of the free parameters, 
\begin{equation} \sum_{i}\frac{\delta ln\Phi_{i}}{\delta\Theta}=0, \end{equation}
a set of nonlinear equations can be defined as the following.

\begin{equation} f : \sum_{i} \frac{\Phi_{f}-\Phi_{c}}{\Phi}=0, \end{equation}

\begin{equation} \sigma : \sum_{i}\frac{\Phi_{c}}{\Phi}\left(\frac{(\mu_{x}-\mu_{xc})^{2}+(\mu_{y}-\mu_{yc})^{2}}{\sigma^{2}}-2\right)= 0, \end{equation}

\begin{equation} \Sigma_{x} : \sum_{i}\frac{\Phi_{f}}{\Phi}\left(\frac{(\mu_{x}-\mu_{xf})^{2}}{\Sigma_{x}^{2}}-1\right)= 0, \end{equation}

\begin{equation} \mu_{xf} : \sum_{i}\frac{\Phi_{f}}{\Phi}(\mu_{x}-\mu_{xf}) = 0, \end{equation}

\begin{equation} \mu_{xc} : \sum_{i}\frac{\Phi_{c}}{\Phi}(\mu_{x}-\mu_{xc}) = 0, \end{equation}

\begin{equation} \mu_{yc} : \sum_{i}\frac{\Phi_{c}}{\Phi}(\mu_{y}-\mu_{yc}) = 0, \end{equation}

\begin{equation} \tau : \sum_{i}\frac{\Phi_{f}}{\Phi}\left(\frac{\mu_{y}}{\tau}-1-c_{o}(\mu_{1}e^{-\frac{\mu_1}{\tau}}-\mu_{2}e^{-\frac{\mu_{2}}{\tau}})\right)= 0. \end{equation}

\begin{table*}
 \centering
  \caption{Fitted parameter values for the set of magnitude bins analysed by the VPD-Probability Fitting Routine.}\label{VPD_Fit}
  \begin{tabular}{@{}cccccccc@{}}
   \hline \hline Interval & f & $\sigma$ & $\mu_{xc}$ & $\mu_{yc}$ &
   $\tau$ & $\Sigma_{x}$ & $\mu_{xf}$ \\

   \hline   12.00 $<$ Z $<$   14.00 &   0.75 &   3.28 &   3.53 &  27.73 &  21.46 &  18.94 &  -5.43 \\
  14.00 $<$ Z $<$   16.00 &   0.71 &   5.45 &   4.43 &  30.65 &  14.26 &  18.89 &  -4.00 \\
  16.00 $<$ Z $<$   18.00 &   0.89 &   5.43 &  -0.27 &  31.87 &  14.91 &  19.54 &  -1.95 \\
 \hline
   \end{tabular}
\end{table*}

\begin{figure}
\begin{center}
\includegraphics[width=\columnwidth]{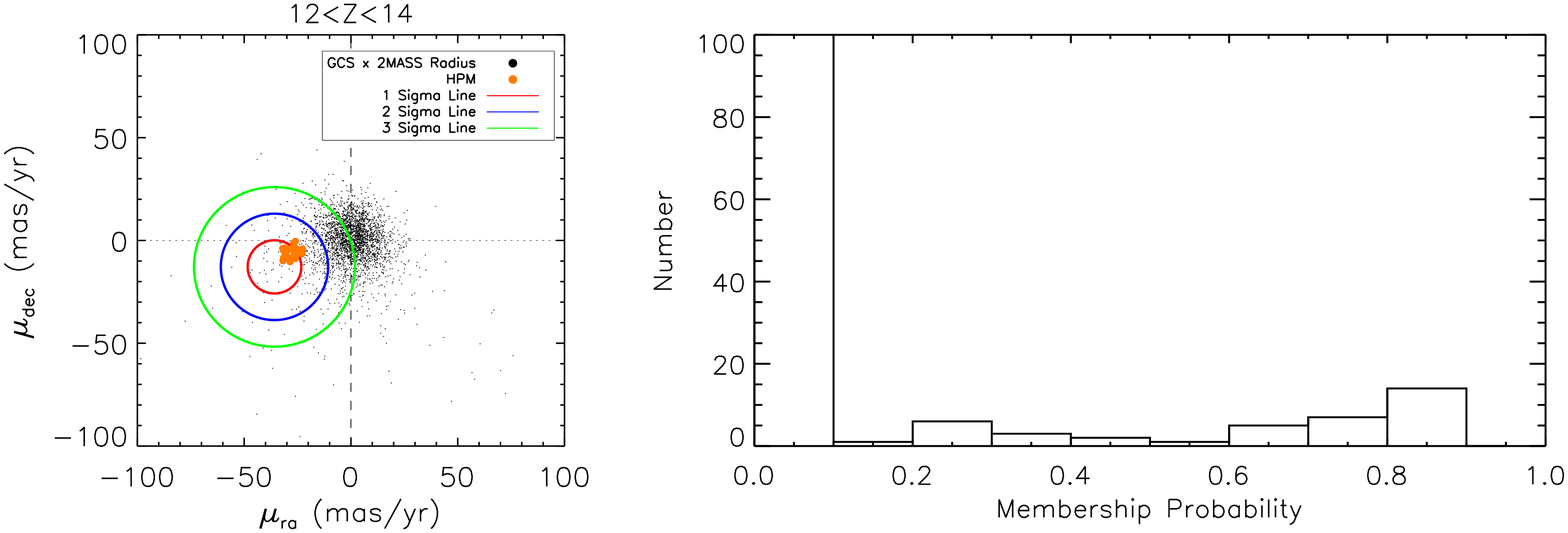}
\includegraphics[width=\columnwidth]{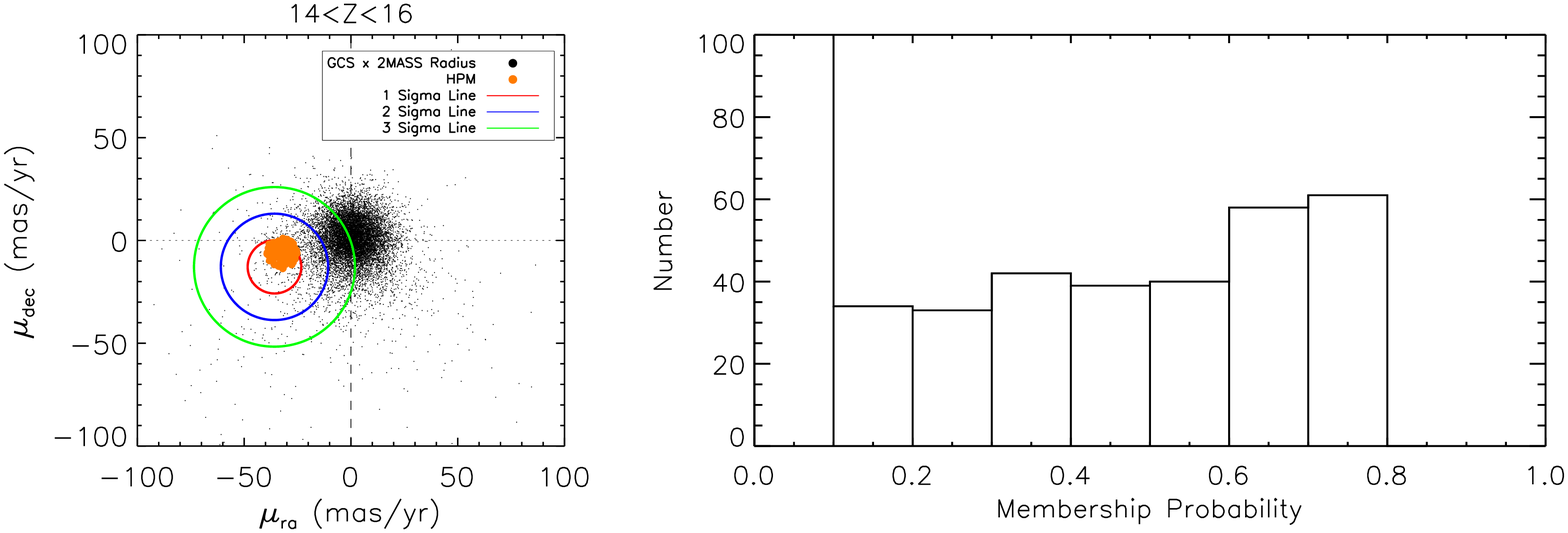}
\includegraphics[width=\columnwidth]{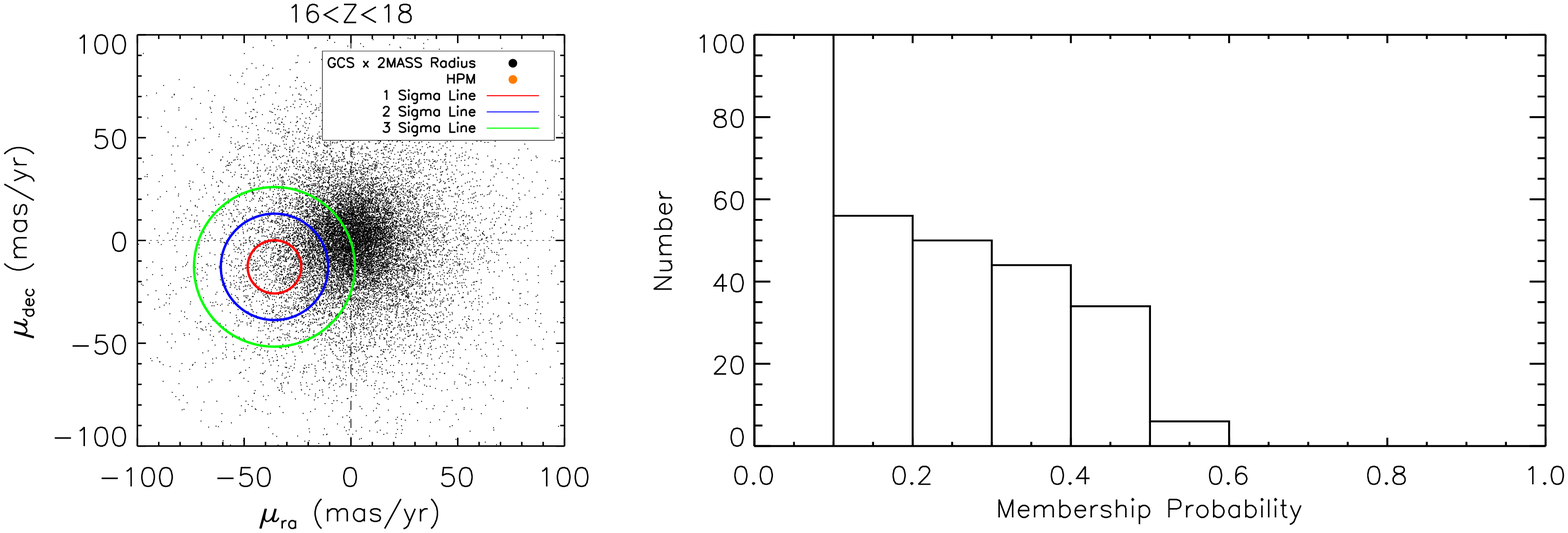}
\caption{Proper Motion vector point diagrams and the resulting
probability histograms for each magnitude interval in the probability
fitting of the Praesepe data. In each plot the three coloured lines
represent the 1$\sigma$, 2$\sigma$ and 3$\sigma$ errors obtained from
our estimate of proper motion errors by way of the 2D Gaussian fit,
they have been shifted to be located over the clusters proper motion
centre. The points in orange show our high probability sample.}\label{pm_hist}
\end{center}
\end{figure}

To each of these equations a bi-section algorithm was devised that
checks for root bracketing and then proceeds to find the root to the
desired level of accuracy. Once one parameter has been fitted, the next
parameter is subject to the same process. Finally once $\tau$ has been
found the process reverts back to its starting point and runs again
until all the free parameters have been fixed (showing no further sign
of deviation). To start the process a set of initial values is
required for the free parameters, these are as in
\citet{Deacon_2004}. Once the values are fixed the membership
probabilities for the $i^{th}$ object can be calculated thus;

\begin{equation} p_{i}=\frac{(1-f)\Phi_{ci}}{\Phi_{i}}. \end{equation}

The fitted values for each magnitude interval are shown in Table~\ref{VPD_Fit}
and vector point diagrams and probability histograms in Figure~\ref{pm_hist}. Taking 
the sigma value of the Gaussian from section 3.1 and placing it over the cluster's 
centre of proper motion we find 380 objects with a probability assignment. Of these, 121 
have probabilities that place them in the High Membership bracket (p $>$ 0.60). In total 
145 sources are found to have a p $\geq$ 0.60. The 24 laying outside the sigma selection are still 
very much consistent with the cluster as can be seen by small spread of objects proper 
motion space in Figure~\ref{pm_hist}.

\section{Members of Praesepe from SDSS}
In this section the process for locating any possible candidates/members of the Praesepe
cluster from available data within the cross link between UKIDSS and SDSS will be described.
\begin{enumerate} 
\item Select those objects that are again within the 3 degree radial cut from 
the cluster centre. These objects are only present in UKIDSS-SDSS data set Any corresponding SDSS-2MASS source would have been 
treated within the previous section.
\item Select the list of candidate objects from 6 CMDs.
\item Extract the FITS files for these objects from the WSA and SDSS Data Access Server.
\item Calculate a pixel to pixel transformation between the two images and using the epoch difference calculate a proper motion.
\item Perform a basic probability analysis on the resulting data.
\end{enumerate}

\subsection{Extracting the Candidates}

\begin{table*}
\suppressfloats[t]
 \centering
  \caption{Probability of membership, magnitude range for our methods of calculating probabilities of membership using the annulus as well as the two control
areas.}\label{SDSS_Prob_table}
  \begin{tabular}{@{}cccccccc@{}}
   \hline \hline Magnitude Range & Probability Annulus & Probability & Probability\\
                  &  & $\mu_\alpha$=0 mas yr$^{-1}$ & $\mu_\alpha$=+35.66 mas yr$^{-1}$\\
                  &  & $\mu_\delta$=+37.85 mas yr$^{-1}$ & $\mu_\delta$=-12.70 mas yr$^{-1}$\\ 

   \hline 15 $<$ Z $<$ 16 &  0.000 &  0.000 &  0.000 \\
16 $<$ Z $<$ 17 &  0.000 &  0.000 &  0.000 \\
17 $<$ Z $<$ 18 &  0.000 &  0.000 &  0.000 \\
18 $<$ Z $<$ 19 &  0.000 &  1.000 &  0.000 \\
19 $<$ Z $<$ 20 &  0.581 &  0.500 &  1.000 \\
20 $<$ Z $<$ 21 &  0.000 &  0.000 &  0.000 \\
 \hline
   \end{tabular}
\end{table*}

\begin{figure*}
\begin{center}
\includegraphics[width=2\columnwidth]{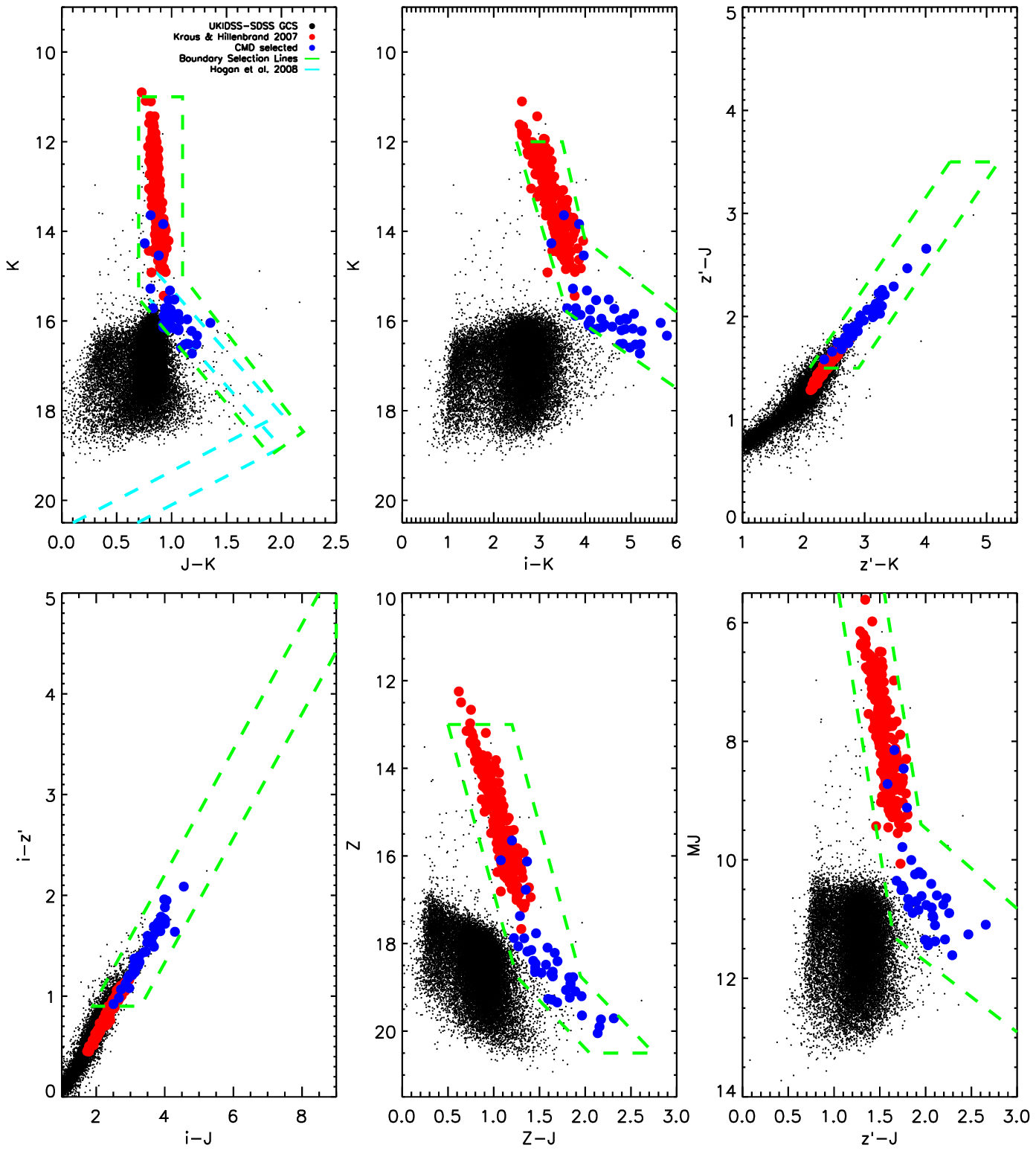}
\caption{CMDs for SDSS and UKIDSS photometric bands showing the selection regions 
(as denoted by the green dashed lines) used to extract the set of 39 candidate objects as shown by the blue circles. The objects of \citet{Kraus_2007} shown in red 
have been used alongside information from plots in figure 9 of \citet{Hawley_2002} to denote the appropriate regions for the cluster 
sequence. Additionally the cluster sequence for the Hyades (a cluster of similar age to Praesepe) as found 
by \citet{Hogan_2008} has been corrected for the distance of Praesepe and over plotted in the $J$-$K$, $K$ plot (blue dashed line) to help trace any possible cluster sequence. The black dots are the UKIDSS-SDSS sources that are not present in the UKIDSS-2MASS population.} 
\label{sdss_selct_CMD}
\end{center}
\end{figure*}

Performing the same radial cut with objects only present in the UKIDSS-SDSS 
match (no 2MASS counterpart) led to a selection of 22,564 objects. To extract the candidates 
from this list, a series of photometric cuts were 
applied in a range of different CMDs. These can be seen in Figure~\ref{sdss_selct_CMD}. 
These CMDs made use of a range of different filters and known locations of LMS/VLMS 
and BDs as detailed by \citet{Hawley_2002}. Objects within these predefined regions were selected for further analysis.

\subsection{Refinement of Proper Motions}
The primary problem with the SDSS dataset is the small epoch 
difference between the two surveys which subsequently results in large errors on the proper motion. This, coupled with 
a small number of objects ($<$ 100) means that the analysis as performed for the 2MASS data is not applicable. 
An alternate approach has therefore been taken in order to assign membership probabilities to these candidates. To begin with, 
suitable queries were developed to allow the acquisition of the catalogue and fits flat files from each survey's data centre
\footnote{For the SDSS data this involved creating a few specific requests as the 
catalogue data is split amongst a wide range of files. The SDSS DAS has also recently been updated to SDSS 
DAS Version 2 for use with DR7, users are instructed to read the DR7 release website for more information.}. 
The candidate object were then located and a set of reference objects 
selected. These reference objects had to be common to both chips (appearing in each epoch), be in the 
magnitude range 12.0 $< Z <$ 18.0, have ellipticities less than 0.2 (from UKIDSS catalogue data) and not sit 
within 5$\%$ of the chip border (in pixels) in order to help minimise any radial distortion effects. 

The list of reference objects are used to create a 12 parameter transformation (where too few reference objects were present for the 
quadratic fit a linear 6 parameter fit was tried instead) that allows the motion of the candidate object between the 
two epochs to be calculated. Reference objects that are shown to be moving at a rate greater than 3 times the rms value of the fit are 
discarded. The revised reference list is again passed to the fitting routines in order to calculate the correct coefficients that describe 
the motion between the two epochs. This motion is then converted into mas yr$^{-1}$ through the use 
of the following equations;

\begin{center}
\begin{equation} \mu_\alpha=\frac{((CD1\_1\times \Delta x+\Delta y\times CD1\_2) \times 3600 \times 1000)}{epoch\_difference}, \end{equation}
\begin{equation} \mu_\delta=\frac{((CD2\_1\times \Delta x+\Delta y\times CD2\_2) \times 3600 \times 1000)}{epoch\_difference}. \end{equation}
\end{center}

The quantities $CD1\_1$, $CD1\_2$, $CD2\_1$ and $CD2\_2$ are the world co-ordinate transformation matrix elements contained within the 
FITS header (\citealt{Greisen_2002}),whilst $\Delta x$ and $\Delta y$ simply refer to the change in pixel elements. The direction in $\mu_\alpha$ is finally further 
corrected by multiplying by the appropriate cos$\delta$ value.

In order to calculate errors, the magnitudes of all possible objects were 
calculated on the reference image frames using the relevant calculations 
listed on the SDSS\footnote{SDSS Magnitude Algorithm calculations: 
\hbox{http://www.sdss.org/dr6/algorithms/fluxcal.html$\#$counts2mag}} and 
WSA\footnote{WFCAM Magnitude Algorithm calculations: \hbox{http://surveys.roe.ac.uk/wsa/flatFiles.html$\#$catmags}} 
pages. The x and y pixel centroiding errors reported for each object in the catalogue 
files were then totalled for each bin of width two magnitudes and divided by the number in that magnitude bin to give an average 
centroiding uncertainty value. These values were then added in quadrature with the rms in that particular direction as found by the 12 parameter quadratic fit.

\begin{center}
\begin{equation} error\_pm=\sqrt{rms^{2}+err_{epoch 1}^{2}+err_{epoch 2}^{2}} \end{equation}
\end{center}
The values representing the error found in the x or y direction for both epoch 1 and epoch 2 measurements are in turn calculated from the following;
\begin{center}
\begin{equation} \mu_\alpha\_err=\frac{((CD1\_1\times xerr+yerr\times CD1\_2)\times3600\times1000)}{epoch\_difference} \end{equation}
\begin{equation} \mu_\delta\_err=\frac{((CD2\_1\times xerr+yerr\times CD2\_2)\times3600\times1000)}{epoch\_difference} \end{equation}
\end{center}

where xerr and yerr represent the centroiding errors in the x and y directions as described in the previous paragraph. 
The errors on the proper motion are likely an overestimation of the astrometic errors as they are based on the rms from the 
fitting process rather than an error in the transformed coordinate.
 
To try and calculate membership probabilities for these objects, an attempt at using control data to determine 
levels of contamination was undertaken.The cluster circle and two control circles of radius 26'' were used. 
The radius value is the average value taken from the 12 parameter fit information, ignoring any 
obvious discrepancies (as objects become fainter the centroiding errors become larger). Those that were in the faintest magnitude bin had 
centroiding errors a factor ten larger than for the other candidates and so were discounted from the average). The circles 
were located at the same distance from (0,0) in proper motion space and as is usual, 
the data were split into magnitude bins with the probability being calculated 
using equation~\ref{prob_eq}.

\begin{center} 
\begin{equation}
P_{membership}=\frac{N_{cluster}-N_{control}}{N_{cluster}}. 
\label{prob_eq}
\end{equation}
\end{center}

\begin{figure}
\begin{center}
\includegraphics[width=\columnwidth]{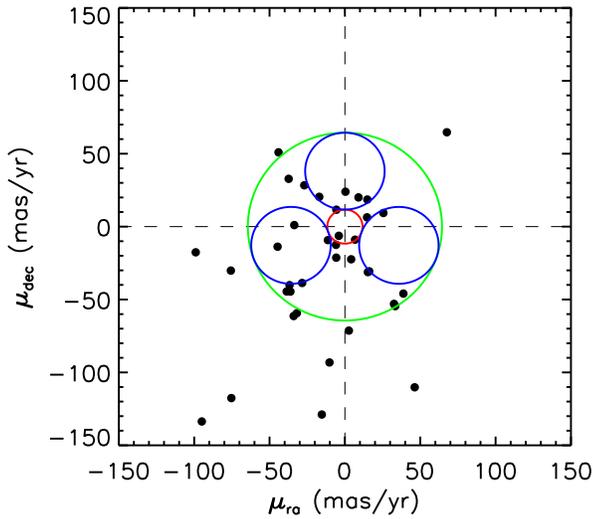}
\caption{Proper motion vector diagram for the 39 UKIDSS-SDSS candidate objects selected from the photometric cuts. The cluster location 
is at $\mu_{\alpha} = -35.81$ mas yr$^{-1}$ \& $\mu_{\delta} = -12.85$ mas yr$^{-1}$ with the two control circles and the annulus 
used for the second measure of probability also plotted. None of these objects are included in the luminosity and mass function analysis.}
\label{SDSS_Prob}
\end{center}
\end{figure}

P$_{membership}$ is the probability assigned for a particular magnitude bin, N$_{cluster}$ the number of field 
and cluster stars within the cluster circle and N$_{control}$ the number of field stars. Thus N$_{cluster}$-N$_{control}$ 
should give the number of Praesepe members. One flaw to this method as reported by \citet{Casewell_2007} is that this 
method is dependent on the chosen location of the control circle. A solution to correct this is to use the field star count 
within the annulus of Figure~\ref{SDSS_Prob} and scale this to the area of the cluster circle to estimate levels of contamination. 
The resulting probabilities for the control circles and annulus are reported in Table~\ref{SDSS_Prob_table}. Any negative probabilities have been altered to 0.00 
The only magnitude bin to contain a positive probability is the bin $19 < Z \leq 20$, for which the control data revealed a probability of 0.5 and 1.0. 
The annulus method for the same magnitude bin put the probability at 0.58. These membership probabilities should be viewed with caution 
as the low numbers (i.e. few objects) make the probabilities difficult to fully substantiate. As such, objects in these fainter magnitude bins have not been 
taking into consideration when constructing the luminosity and mass functions reported in section~\ref{CMLF}. When 
the survey of the cluster centre is complete it is hoped that not only will more LMS/BD candidates be found but that the increase in numbers will also 
allow for a more robust membership probability to be assigned.  

\section{Cluster Mass and Luminosity Functions}
\label{CMLF}
\begin{figure}
\begin{center}
\includegraphics[width=\columnwidth]{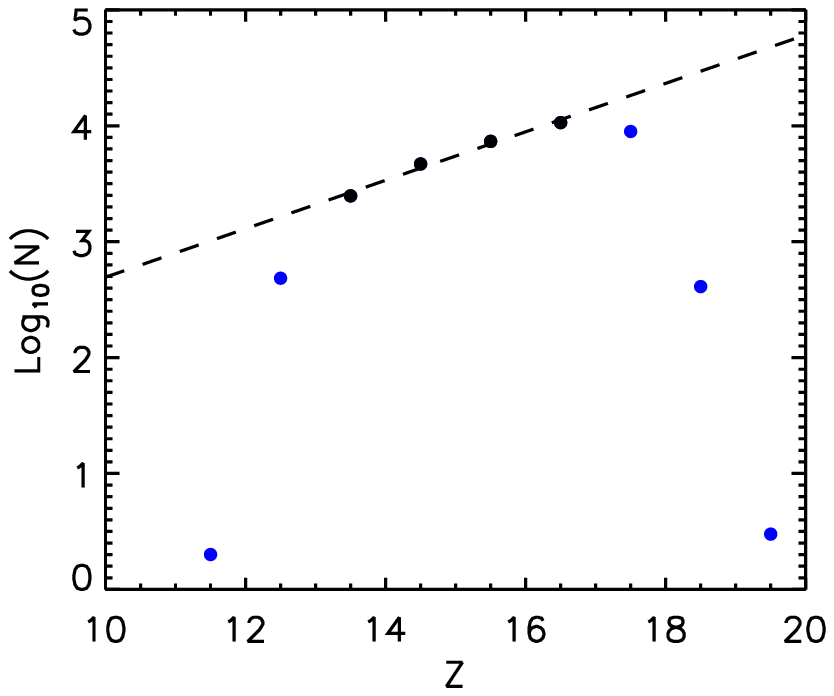}
\includegraphics[width=\columnwidth]{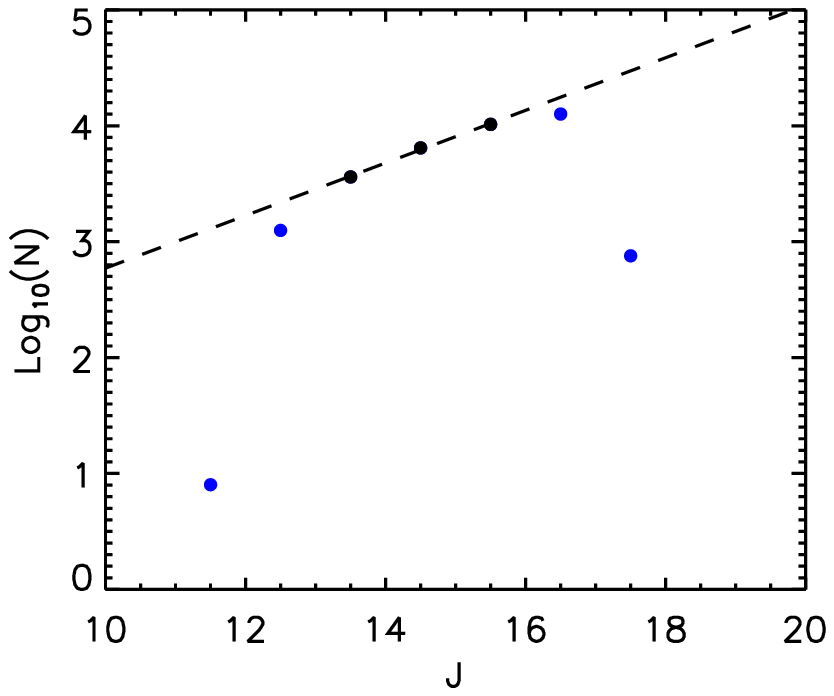}
\includegraphics[width=\columnwidth]{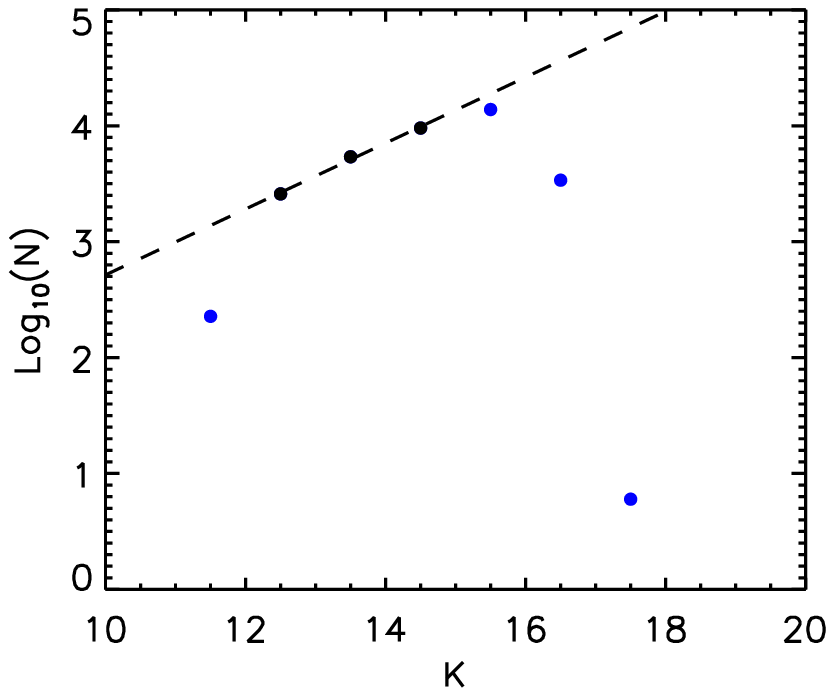}
\caption{An estimate of the incompleteness in the 2MASS selected Praesepe data set for the UKIDSS $Z$, $J$ and $K$ bands respectively. The black dotted line
is a least squares fit to the associated black points in figure. There is clearly a drop off at the end of the functions whilst the deficit for the 
brightest magnitude bins is caused by saturation effects.}
\label{incompleteness_2MASS}
\end{center}
\end{figure}

\begin{figure}
\begin{center}
\includegraphics[width=\columnwidth]{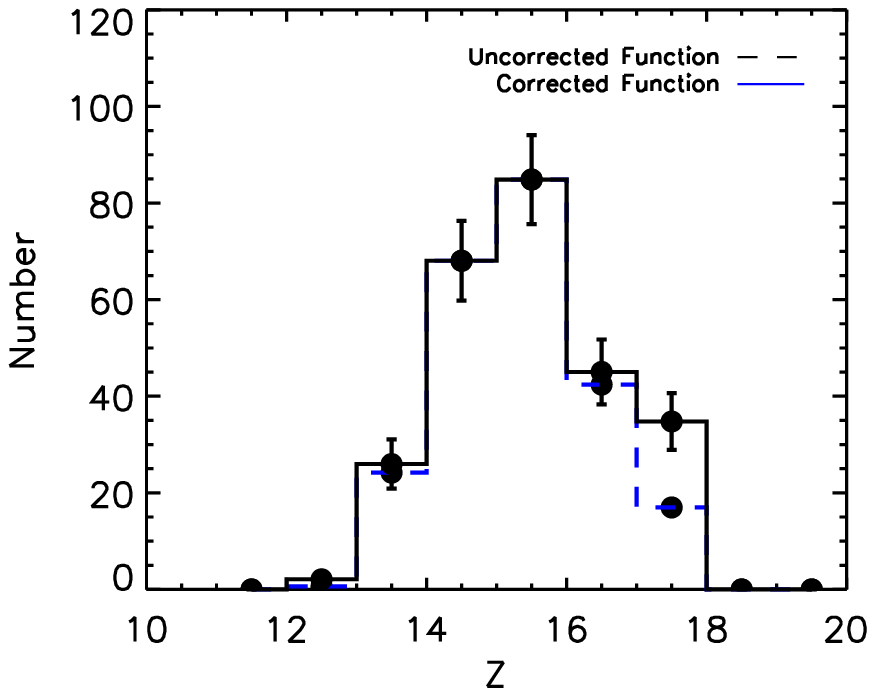}
\includegraphics[width=\columnwidth]{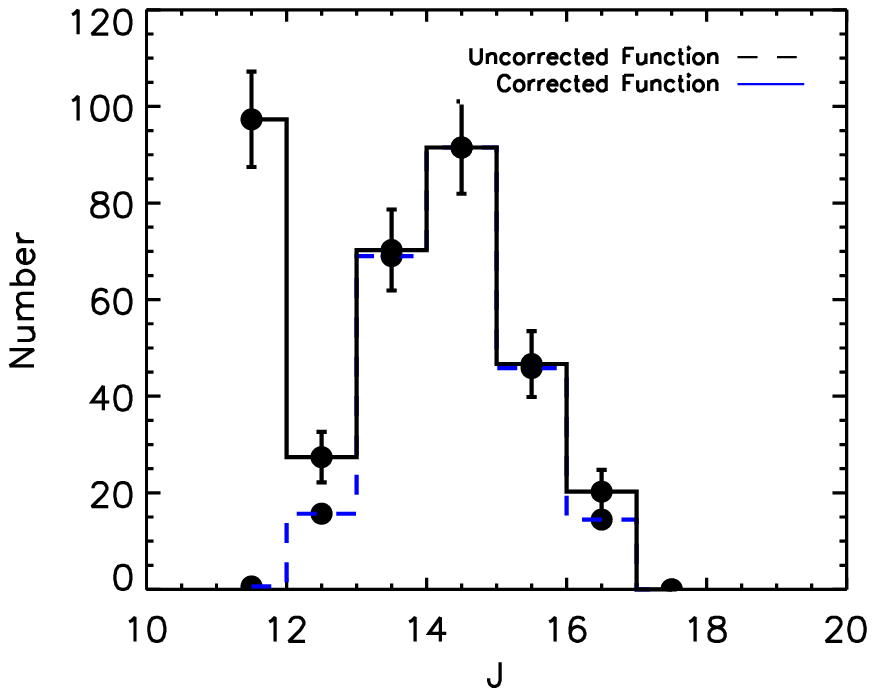}
\includegraphics[width=\columnwidth]{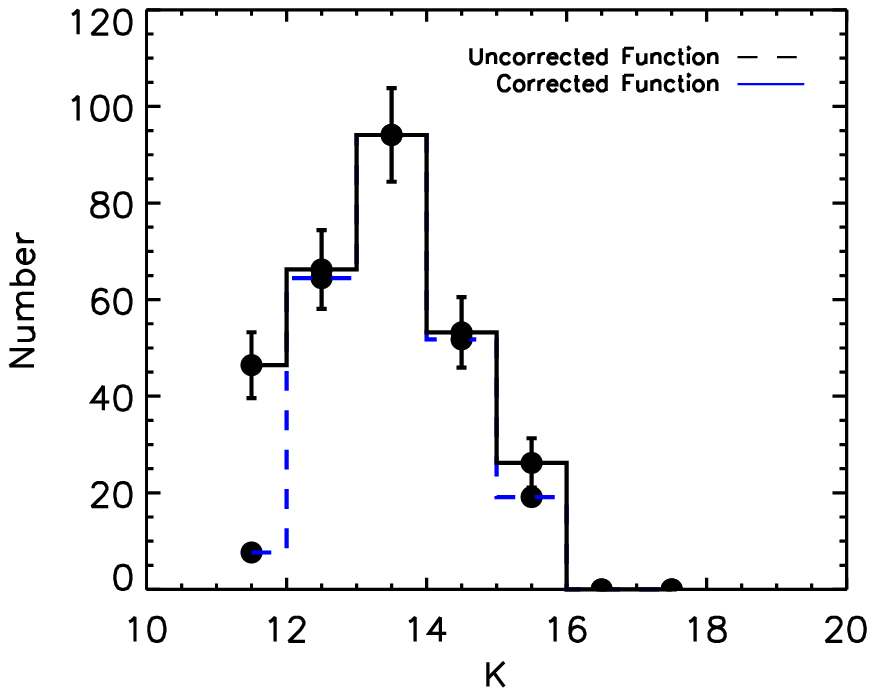}
\caption{ The Luminosity Functions derived for the Praesepe star cluster in the $Z$, $J$ and $K$ bands. The distribution rises to a peak 
before starting to decay due to the change in the slope of the mass-luminosity relationship. The error bars are Poisson errors.}
\label{l_f}
\end{center}
\end{figure}

In order to produce the luminosity function of the cluster the effects of incompleteness on the data must be taken into account. 
As magnitude increases, the number of objects in a specified magnitude bin of uniform
distribution will also increase. Counting the number of stars present and comparing it to this predicted rate of 
growth will show up any signs of incompleteness. This is done by taking the logarithm of the number of stars in the magnitude bin and fitting 
a best fit line to the data up until this ``drop off'' point. An estimate for the level of incompleteness is calculated from the deficit. 
For the purposes of this study the best fit line was fitted between the black points in 
Figure~\ref{incompleteness_2MASS} for the UKIDSS-2MASS data set in the $Z$, $J$ and $K$ filters.  The drop off can be clearly seen at the
fainter magnitudes with the brighter magnitude drop off likely caused
by bright cut off limit of the surveys.

The cluster luminosity function (Figure~\ref{l_f}) was calculated by summing the
assigned membership probabilities of each object in bins of width one
magnitude. Each interval was then multiplied by the incompleteness
factor, producing the correction from the dashed blue to solid
black lines. In order to avoid distorting the luminosity function we applied this process 
to the UKIDSS-2MASS population only. The fainter magnitudes and tentative membership probabilities 
of the UKIDSS-SDSS candidates excluded them, as inappropriately large correction factors would have to be applied. 
The luminosity function can be seen to have a clear and 
well defined peak in each of the filters followed by a decrease due to a change in slope of the mass-luminosity relationship. 
The first point in the $J$-band luminosity function has been artificially raised due to an 
large incompleteness factor being applied. This point is thus not treated when fitting the mass function.
The errors shown are simply Poisson errors.

\begin{figure}
\begin{center}
\includegraphics[width=\columnwidth]{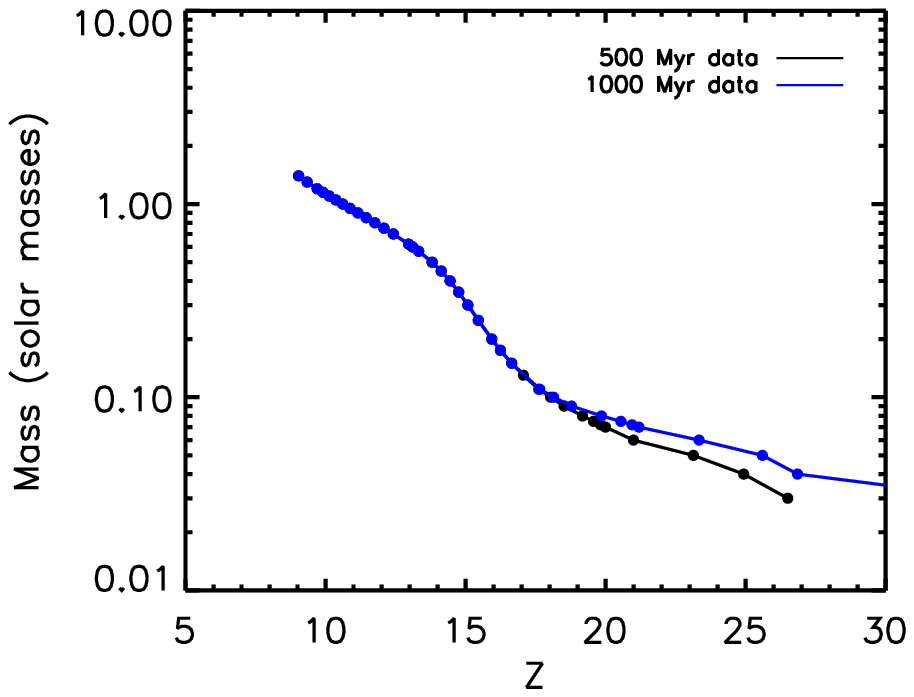}
\includegraphics[width=\columnwidth]{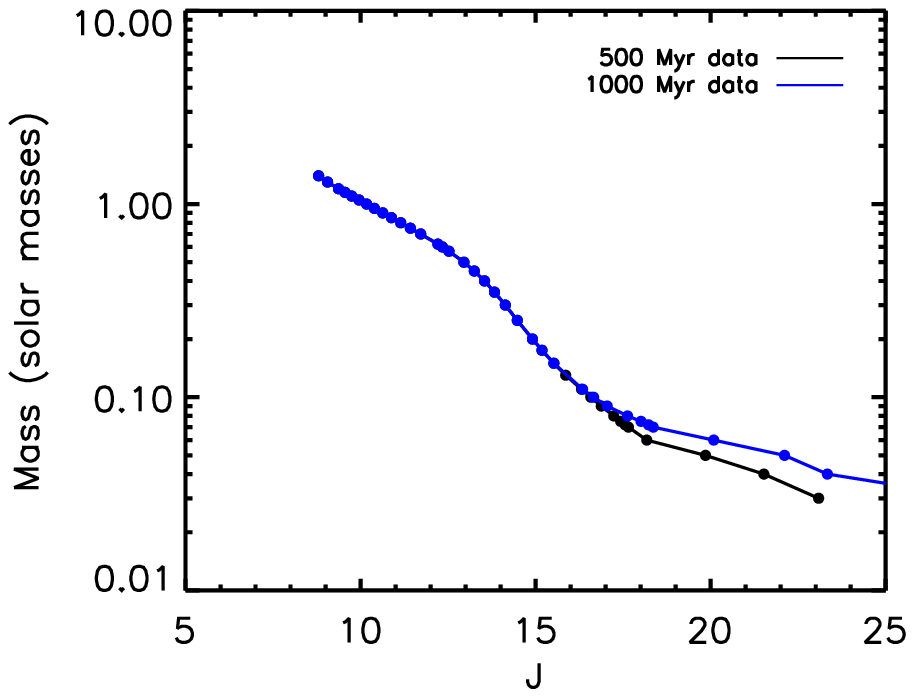}
\includegraphics[width=\columnwidth]{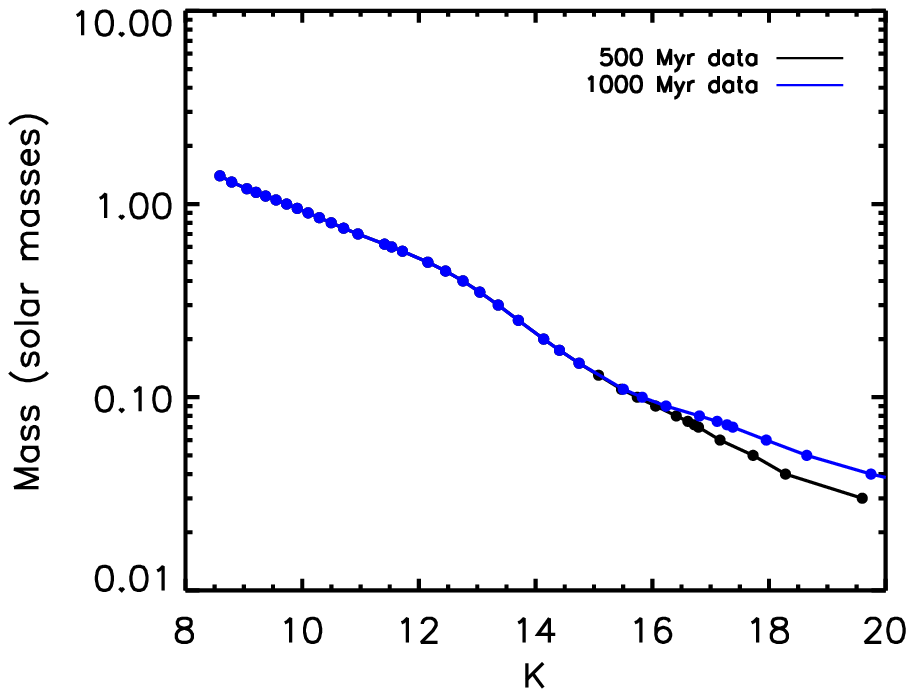}
\caption{The mass-luminosity relationship used in this study is based off
a composite BT-NextGen and BT-DUSTY line with a cubic spline interpolation. 
The black points represent the 500 Myr data whilst the blue
the 1 Gyr data. The lines only start to deviate appreciably at faint magnitudes, which we do not investigate. 
Shown here is the mass-luminosity relationship for the $Z$, $J$ and $K$ bands respectively.}\label{m_l_relation}
\end{center}
\end{figure}

To convert the cluster luminosity function into a mass function a
mass-luminosity relationship is required (see Figure~\ref{m_l_relation}). Again a
combined relationship formed from the BT-NextGen and BT-DUSTY models was
constructed. Cubic spline interpolation was used to interpolate between the points. 
As the age of the Praesepe cluster is not well confined this mass-luminosity
relationship has again been constructed based of both the 500 Myr and
1 Gyr model data.

If one considers a magnitude interval of width dM the total number of
stars (dN) will be given by

\begin{equation} dN=\Phi(M)dM \end{equation}

where $\Phi$(M) is the luminosity function. In the corresponding mass
interval the same dN is now given by $\xi$(m)dm, where dm is the
interval in the mass range and $\xi$(m) is the mass function.

Hence to calculate the mass function the previous equations can be
easily rearranged to the form of

\begin{center} \begin{equation} \xi(m)=\frac{\Phi(M)dM}{dm}
\end{equation} 
\end{center}

\begin{figure}
\begin{center}
\includegraphics[width=\columnwidth]{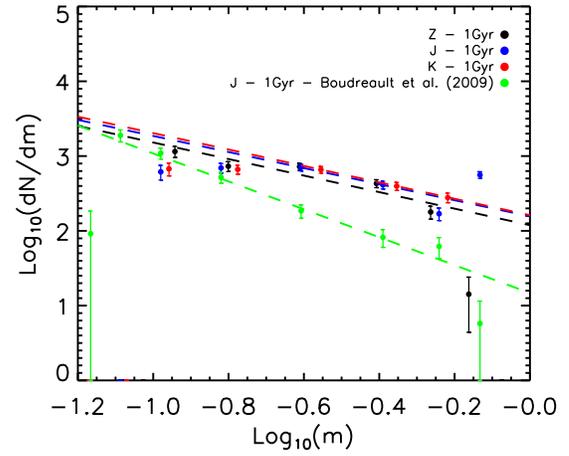}
\caption{The derived cluster mass function with the power law fit for the $Z$ photometric band shown in dashed black. 
The fits for the $J$ and $K$ bands are also shown in blue and red respectively. 
The green points are taken from \citet{Boudreault_2009}. The error bars are Poisson errors. 
For clarity only the 1 Gyr data and fit has been plotted as there is little deviation between 
that and the 500 Myr fit.}
\label{mass_function}
\end{center}
\end{figure}

In Figure~\ref{mass_function}, Praesepe's cluster mass function is plotted with a single power law 
fit for the points between 0.6 and 0.125 \msun for the $Z$ photometric band. The $J$ and $K$ bands have been treated similarly but due to the corrections 
applied for the incompleteness affecting a wider range of magnitudes a narrower mass range was considered.
The form of the power law is shown below.
\begin{equation}
\xi(m) \propto m^{-\alpha}
\end{equation}

The results of this fitting are that $\alpha_{500Myr}$=1.11 $\pm$ 0.37 and $\alpha_{1Gyr}$=1.10 $\pm$ 0.37.
For the $J$ and $K$ bands the results were  $\alpha_{500Myr}$=1.07, $\alpha_{1Gyr}$=1.07 and $\alpha_{500Myr}$=1.09, $\alpha_{1Gyr}$=1.09 respectively.
These values are much lower than the Salpeter IMF (2.35) but the upper limits agree roughly with the values calculated and 
cited by \citet{Kraus_2007} of $\alpha$=1.4 $\pm$ 0.2. The \citeauthor{Kraus_2007} result was derived using a 
mass-spectral type relationship, however after retrieving 2MASS $J$ band photometry for their objects and constructing 
the $J$ band luminosity function as we have done for our UKIDSS-2MASS sample we find $\alpha$ to differ by only 0.1.

\begin{table*}
 \centering
  \caption{Mass function results from previous surveys}\label{Mass_function_table}
  \begin{tabular}{@{}ccccc@{}}
   \hline \hline Survey & Passband & Mass range & Slope\\
   \hline
                 Baker et al. & $Z$ & 0.125 - 0.6 \msun & 1.10 $\pm$ 0.37\\
		 Baker et al. & $J$ & 0.20  - 0.5 \msun & 1.07 \\
		 Baker et al. & $K$ & 0.20  - 0.5 \msun & 1.09 \\
		 Kraus \& Hillenbrand & SED$^{a}$ & 0.17 - 1 \msun & 1.4 $\pm$ 0.2\\
		 Boudreault et al. & $J$ & 0.10  - 0.7 \msun & 2.3 $\pm$ 0.2\\
		 Boudreault et al. & $J$ & 0.18  - 0.45 \msun & 1.8 $\pm$ 0.1\\
   \hline 
   \end{tabular}
   \begin{tabular}{@{}c@{}}
     $^{a}$ See the appendix of Kraus \& Hillenbrand (2007) for a discussion on how the multiple photometric bands \\
     coupled with theoretical models were used to derive masses for each spectral type.
   \end{tabular}
\end{table*}

 A more recent survey of the cluster centre has also just been carried 
out by \cite{Boudreault_2009}. The authors identify some 150 candidate members with 6 expected to be BDs. Of these 6 only 3 are currently 
within our survey region, objects 55, 909 and 910. Objects 55 and 909 are precluded from our search due to their morphological classification in the 
UKIDSS data. Object 910 is present and photometrically appears to agree with \citeauthor{Boudreault_2009} ($J=17.66$, $K=16.8$). The presence of 
other photometric bands has however shown this object to be far too blue in the $Z$-$J$,$J$ diagram ($Z=18.33$) for it to be considered as a 
cluster member by our current analysis. The value presented by \citeauthor{Boudreault_2009} of $\alpha = 1.8 \pm 0.1$ for $\xi(m)$ (the mass function) 
appears to be much greater than our calculated value and more in line with the upper value presented by \citeauthor{Kraus_2007}. 
The survey however does not use proper motion information and restricts its objects to candidates with uniform probability of 
membership (p=1.0). Obtaining the $J$ band photometry of the 150 \citeauthor{Boudreault_2009} objects and treating them in the 
same manner again with our luminosity-mass relationship confirms their result as we find $\alpha = 1.85 \pm 0.15$. 
See Table~\ref{Mass_function_table} for a summary of the results. It is important then that a full survey including the 
more densely populated centre takes place as this would likely increase the number of members 
found (particularly at the lower mass end) thus altering the IMF that would be observed. The full extent of this change is of course unknown. 
It is clear however that even with a more complete data set there are likely to be few BDs within the 
Praesepe cluster. This is consistent with a cluster of such age having undergone many cycles of equipartition of energy, so ejecting low mass 
objects from the cluster. The true record breakers in the low mass stakes will therefore be found as companions to other objects as the combined 
system mass would restrict the amount ejected through dynamical events.

\begin{figure}
\begin{center}
\includegraphics[width=\columnwidth]{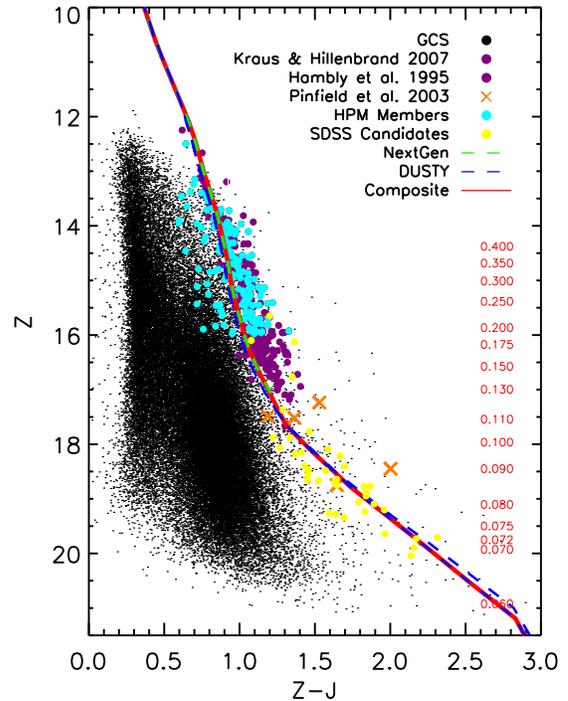}
\caption{A $Z$-$J$,$Z$ CMD showing the cluster members given by \citeauthor{Hambly_1995a} and \citeauthor{Kraus_2007} 
alongside the 145 HPM derived from this work. The BT-NextGen, BT-DUSTY and composite selection lines are also shown as are 
the masses (in units of solar mass) given by those models for the 500 Myr case. The candidate UKIDSS-SDSS members are also shown to show the 
limits of the survey. Most of the UKIDSS-SDSS objects were found in this analysis to be non-members.}
\label{CMD_final}
\end{center}
\end{figure}

\section{Summary}
A study of available UKIDSS data on the Praesepe star cluster combined with archive data from the 2MASS and SDSS surveys 
for the range of 12 $\leq Z <$ 21 has found, through a combination of proper motion and colour magnitude diagram 
selections 145 HPM (see Figure~\ref{CMD_final} and Appendix ~\ref{HPM_Appendix}), of which, 14 appear to be new members, Appendix ~\ref{HPM_new_Appendix}. 
The majority of the detected HPM objects have also been found in more than one previous body of work, almost certainly 
confirming their status as cluster members. These objects all inhabit a fairly bright magnitude region on the cluster 
sequence as they have been derived from investigation the brighter UKIDSS-2MASS data set. 
The UKIDSS-SDSS data set, whilst allowing fainter objects to be examined, suffers from a short time base 
between observations and small number statistics making membership assignments difficult to quantify 
(see Appendix~\ref{SDSS_cands_appendix} for a list of the candidate objects). 
An investigation into the cluster luminosity and hence mass function 
was also carried out. The upper limits of the later agree with a previous result given by \citet{Kraus_2007}, confirming that although 
a low mass population of objects may have existed the dynamical evolution of the cluster over its lifetime has led to 
a significant population depletion. Only when a full survey of the cluster with UKIDSS is complete and second epoch $K$ data 
available can one hope to truly evaluate this cluster.

\section{Acknowledgements}
DEAB acknowledges the support of an STFC Postgraduate studentship. NL was supported by the Ram\'on y Cajal fellowship number 08-303-01-02. 
This publication makes use of data products from the United Kingdom Infrared Deep Sky Survey, the Sloan Digital Sky Survey, 
the Two Micron All Sky Survey, the SIMBAD database, operated at CDS, 
Strasbourg, France and NASA's Astrophyics Data System Bibliographic Services. The SDSS is managed by the 
Astrophysical Research Consortium (ARC) for the Participating Institutions. The Participating Institutions are 
The University of Chicago, Fermilab, the Institute for Advanced Study, the Japan Participation Group, The Johns Hopkins 
University, Los Alamos National Laboratory, the Max-Planck-Institute for Astronomy (MPIA), the Max-Planck-Institute for 
Astrophysics (MPA), New Mexico State University, University of Pittsburgh, Princeton University, the United States Naval 
Observatory, and the University of Washington. Funding for the Sloan Digital Sky Survey (SDSS) has been provided by the 
Alfred P. Sloan Foundation, the Participating Institutions, the National Aeronautics and Space Administration, the National 
Science Foundation, the U.S. Department of Energy, the Japanese Monbukagakusho, and the Max Planck Society. 
The SDSS Web site is http://www.sdss.org/. The Two Micron All Sky Survey, is a joint project of the University of 
Massachusetts and the Infrared Processing and Analysis Center/California Institute of Technology, funded by the National 
Aeronautics and Space Administration and the National Science Foundation. The authors would like to thank France Allard for the use of the Phoenix
web simulator at http://phoenix.ens-lyon.fr/simulator/index.faces which
has been used in this research.

\bibliographystyle{mn2e}
\bibliography{Bibliography}

\appendix 

\onecolumn
\section{SQL queries}
\label{SQL_query}
\par
\subsection{UKIDSS-SDSS}
SELECT\\
g.sourceID as u$\_$id,\\
T2.slaveObjID as s$\_$id,\\
g.ra as u$\_$ra,\\
g.dec as u$\_$dec,\\
T2.ra as s$\_$ra,\\
T2.dec as s$\_$dec,\\
m.mjdObs as u$\_$mjd,\\
T2. mjd$\_$z as s$\_$mjd,\\
g.mergedClass as u$\_$class,\\
g.zaperMag3 as u$\_$z,\\
g.zaperMag3Err as u$\_$zerr,\\
g.yaperMag3 as u$\_$y ,\\
g.yaperMag3Err as u$\_$yerr,\\
g.japerMag3 as u$\_$j,\\
g.japerMag3Err as u$\_$jerr,\\
g.haperMag3 as u$\_$h,\\
g.haperMag3Err as u$\_$herr,\\
g.k$\_$1aperMag3 as u$\_$k,\\
g.k$\_$1aperMag3Err as u$\_$kerr,\\
T2.psfMag$\_$u as s$\_$u,\\
T2.psfMagErr$\_$u as s$\_$uerr,\\
T2.psfMag$\_$g as s$\_$g,\\
T2.psfMagErr$\_$g as s$\_$gerr,\\
T2.psfMag$\_$r as s$\_$r,\\
T2.psfMagErr$\_$r as s$\_$rerr,\\
T2.psfMag$\_$i as s$\_$i,\\
T2.psfMagErr$\_$i as s$\_$ierr,\\
T2.psfMag$\_$z as s$\_$z,\\
T2.psfMagErr$\_$z as s$\_$zerr,\\
\\
(T2.distanceMins * 60) as us$\_$separation,\\
\\
3.6e6*COS(RADIANS(g.dec))*(g.ra-T2.ra)/((m.mjdObs-T2.mjd$\_$z)/365.25) AS us$\_$pmra,\\
3.6e6*(g.dec-T2.dec)/((m.mjdObs-T2.mjd$\_$z)/365.25) AS us$\_$pmdec,\\
\\
g.framesetid as framesetid,\\
f.multiframeid as multiframeid,\\
m.filename as filename,\\
m.catname as catname,\\
f.extNum as extnum,\\
T2.run as run,\\
T2.rerun as rerun,\\
T2.camcol as camcol,\\
T2.field as field,\\
T2.rowc$\_$z as xpix,\\
T2.colc$\_$z as ypix,\\
T2.segmentID as segmentID,\\
T2.stripe as stripe,\\
T2.chunkID as chunkID,\\
T2.startmu as startmu\\
\\
FROM\\
gcsMergeLog as ml,\\
Multiframe as m,\\
gcsFrameSets as f,\\
\indent(\\
\indent\indent    SELECT sdss.ra, sdss.dec, x.slaveObjID, x.masterObjID, x.distanceMins,\\
\indent \indent           sdss.psfMag$\_$u, sdss.psfMagErr$\_$u, sdss.psfMag$\_$g, sdss.psfMagErr$\_$g, sdss.psfMag$\_$r, sdss.psfMagErr$\_$r,\\
\indent\indent            sdss.psfMag$\_$i, sdss.psfMagErr$\_$i, sdss.psfMag$\_$z, sdss.psfMagErr$\_$z, f.mjd$\_$z, sdss.run, sdss.rerun,\\
\indent\indent            sdss.camcol, sdss.field, sdss.rowc, sdss.colc,sg.segmentID, sg.stripe, c.chunkID, c.startmu\\
\par
\indent\indent     FROM gcsSourceXDR7PhotoObj as x, BestDR7..PhotoObj as sdss,\\
\indent\indent            BestDR7..Field as f, BestDR7..Segment as sg, BestDR7..Chunk as c\\
\par
\indent\indent     WHERE sdss.type = 6 AND x.slaveObjID = sdss.objID AND f.fieldID = sdss.fieldID \\
\indent\indent     AND f.segmentID = sg.segmentID AND sg.chunkID = c.chunkID\\
\par
\indent\indent\indent          /* Detected in BINNED 1 */\\
\indent\indent\indent	  AND ((flags$\_$i \& 0x10000000) != 0)\\ 
\indent\indent\indent	  AND ((flags$\_$z \& 0x10000000) != 0)\\ 
\indent\indent\indent	  /* Not EDGE, NOPROFILE, PEAKCENTER, NOTCHECKED\\
\indent\indent\indent             PSF$\_$FLUX$\_$INTERP, SATURATED, or BAD$\_$COUNTS$\_$ERROR */\\
\indent\indent\indent	  AND ((flags$\_$i \& 0x8100000c00a4) = 0)\\
\indent\indent\indent	  AND ((flags$\_$z \& 0x8100000c00a4) = 0)\\ 
\indent\indent\indent	  /* Not DEBLEND$\_$NOPEAK or small PSF error */\\
\indent\indent\indent	  AND (((flags$\_$i \& 0x400000000000) = 0) or (psfmagerr$\_$i $<$= 0.2))\\ 
\indent\indent\indent	  AND (((flags$\_$z \& 0x400000000000) = 0) or (psfmagerr$\_$z $<$= 0.2))\\ 
\indent\indent\indent	  /* Not INTERP$\_$CENTER or not COSMIC$\_$RAY */\\
\indent\indent\indent	  AND (((flags$\_$i \& 0x100000000000) = 0) or (flags$\_$i \& 0x1000) = 0)\\ 
\indent\indent\indent	  AND (((flags$\_$z \& 0x100000000000) = 0) or (flags$\_$z \& 0x1000) = 0)\\ 
\par
\indent\indent\indent           AND distanceMins IN (\\
\indent\indent\indent       SELECT MIN(distanceMins) FROM gcsSourceXDR7PhotoObj WHERE\\
\indent\indent\indent              masterObjID = x.masterObjID AND distanceMins $<$ 1.0 / 60\\
\indent\indent      )\\
\indent )\\
\\
AS T2 RIGHT OUTER JOIN gcsSource AS g on g.sourceID=T2.masterObjID\\
WHERE\\
$/$*    Sample selection predicates: Praesepe  RA=120-150 deg \&\& dec=15-25 deg */\\
g.ra BETWEEN 120.0 AND 135.0\\
AND g.dec BETWEEN 15.0 AND 25.0 and\\
(zXi BETWEEN -1.0 AND +1.0 OR zXi $<$ -0.9e9)\\
AND yXi BETWEEN -1.0 AND +1.0\\
AND jXi BETWEEN -1.0 AND +1.0\\
AND hXi BETWEEN -1.0 AND +1.0\\
AND k$\_$1Xi BETWEEN -1.0 AND +1.0\\
AND (zEta BETWEEN -1.0 AND +1.0 OR zEta $<$ -0.9e9)\\
AND yEta BETWEEN -1.0 AND +1.0\\
AND jEta BETWEEN -1.0 AND +1.0\\
AND hEta BETWEEN -1.0 AND +1.0\\
AND k$\_$1Eta BETWEEN -1.0 AND +1.0\\
AND (zClass BETWEEN -2 AND -1 OR zClass $<$ -9999)\\
AND yClass BETWEEN -2 AND -1\\
AND jClass BETWEEN -2 AND -1\\
AND hClass BETWEEN -2 AND -1\\
AND k$\_$1Class BETWEEN -2 AND -1\\
AND (priOrSec = 0 OR priOrSec = g.frameSetID)\\
AND g.frameSetID = ml.frameSetID\\
AND ml.zmfID = m.multiframeID\\
AND g.zppErrBits $<$ 16\\
AND g.yppErrBits $<$ 16\\
AND g.jppErrBits $<$ 16\\
AND g.hppErrBits $<$ 16\\
AND g.k$\_$1ppErrBits $<$ 16\\
AND g.framesetID=f.framesetID\\
AND f.multiframeID=m.multiframeID\\

\subsection{UKIDSS-2MASS}
SELECT\\
g.sourceID as u$\_$id,\\
T2.pts$\_$key as t$\_$id,\\
T2.designation t$\_$designation,\\
g.ra as u$\_$ra,\\
g.dec as u$\_$dec,\\
T2.ra as t$\_$ra,\\
T2.dec as t$\_$dec,\\
m.mjdObs as u$\_$mjd,\\
(T2.jdate-2400000.5) as t$\_$mjd,\\
g.mergedClass as u$\_$class,\\
g.zaperMag3 as  u$\_$z,\\
g.zaperMag3Err as u$\_$zerr,\\
g.yaperMag3 as u$\_$y,\\
g.yaperMag3Err as u$\_$yerr,\\
g.japerMag3 as u$\_$j,\\
g.japerMag3Err as u$\_$jerr,\\
g.haperMag3 as u$\_$h,\\
g.haperMag3Err as u$\_$herr,\\
g.k$\_$1aperMag3 as u$\_$k,\\
g.k$\_$1aperMag3Err as u$\_$kerr,\\
T2.j$\_$m as t$\_$j,\\
T2.h$\_$m as t$\_$h,\\
T2.k$\_$m as t$\_$k,\\
T2.ph$\_$qual as t$\_$phqual,\\
T2.nopt$\_$mchs as nopt$\_$mchs,\\
\\
(T2.distanceMins * 60) as ut$\_$separation,\\
\\
3.6e6*COS(RADIANS(g.dec))*(g.ra-T2.ra)/((m.mjdObs-(T2.jdate-2400000.5))/365.25) AS ut$\_$pmra,\\
3.6e6*(g.dec-T2.dec)/((m.mjdObs-(T2.jdate-2400000.5))/365.25) AS ut$\_$pmdec,\\ 
\\
g.framesetid as framesetid,\\
f.multiframeid as multiframeid,\\
m.filename as filename,\\
m.catname as catname,\\
f.extNum as extnum\\
\\
FROM\\
gcsMergeLog AS ml,\\
Multiframe AS m,\\
gcsFrameSets as f,\\
\par
\indent(\\
\indent\indent    SELECT mass.designation, mass.pts$\_$key, mass.ra, mass.dec, x.slaveObjID, x.masterObjID,\\
\indent\indent    x.distanceMins, mass.jdate, mass.j$\_$m, mass.h$\_$m, mass.k$\_$m,\\
\indent\indent    mass.ph$\_$qual,mass.nopt$\_$mchs\\
\\
\indent\indent    FROM gcsSourceXtwomass$\_$psc AS x, TWOMASS..twomass$\_$psc as mass\\
\\           
\indent\indent    WHERE x.slaveObjID = mass.pts$\_$key AND mass.j$\_$m $>$ 9 AND mass.h$\_$m $>$ 8.5 AND mass.k$\_$m $>$ 8 AND\\
\indent\indent	   distanceMins IN(\\ 
\indent\indent    SELECT MIN(distanceMins) FROM gcsSourceXtwomass$\_$psc WHERE\\
\indent\indent              masterObjID = x.masterObjID AND distanceMins $<$ 1.0 / 60\\
\indent    )\\
)\\
AS T2 RIGHT OUTER JOIN gcsSource AS g on g.sourceID=T2.masterObjID\\
WHERE\\
/$*$    Sample selection predicates:Praesepe  RA=120-135 deg \&\& dec=15-25 deg $*$/\\
g.ra BETWEEN 120.0 AND 135.0\\
AND g.dec BETWEEN 15.0 AND 25.0 and\\
(zXi BETWEEN -1.0 AND +1.0 OR zXi $<$ -0.9e9)\\
AND yXi BETWEEN -1.0 AND +1.0\\
AND jXi BETWEEN -1.0 AND +1.0\\
AND hXi BETWEEN -1.0 AND +1.0\\
AND k$\_$1Xi BETWEEN -1.0 AND +1.0\\
AND (zEta BETWEEN -1.0 AND +1.0 OR zEta $<$ -0.9e9)\\
AND yEta BETWEEN -1.0 AND +1.0\\
AND jEta BETWEEN -1.0 AND +1.0\\
AND hEta BETWEEN -1.0 AND +1.0\\
AND k$\_$1Eta BETWEEN -1.0 AND +1.0\\
AND (zClass BETWEEN -2 AND -1 OR zClass $<$ -9999)\\
AND yClass BETWEEN -2 AND -1\\
AND jClass BETWEEN -2 AND -1\\
AND hClass BETWEEN -2 AND -1\\
AND k$\_$1Class BETWEEN -2 AND -1\\
AND (priOrSec = 0 OR priOrSec = g.frameSetID)\\
AND g.frameSetID = ml.frameSetID\\
AND ml.zmfID = m.multiframeID\\
AND g.zppErrBits $<$ 16\\
AND g.yppErrBits $<$ 16\\
AND g.jppErrBits $<$ 16\\
AND g.hppErrBits $<$ 16\\
AND g.k$\_$1ppErrBits $<$ 16\\
AND g.framesetID=f.framesetID\\
AND f.multiframeID=m.multiframeID\\

\onecolumn
\section{High Probability Members Matched to Previous Works}\footnote{AD = \citet{Adams_2002}, HSHJ = \citet{Hambly_1995b}, JS = \citet{Jones_1983} and KH = \citet{Kraus_2007},.}
\label{HPM_Appendix}
\scriptsize
\begin{center}
\begin{longtable}{cccccccccl}
\caption{High Probability Members Matched to Previous Works}\\
\hline \hline \multicolumn{1}{c}{ID} &
\multicolumn{1}{c}{$Z$} & \multicolumn{1}{c}{$Y$} &
\multicolumn{1}{c}{$J$} & \multicolumn{1}{c}{$H$} &
\multicolumn{1}{c}{$K$} & \multicolumn{1}{c}{$\mu_\alpha$} &
\multicolumn{1}{c}{$\mu_\delta$} & \multicolumn{1}{c}{Pmem} &
\multicolumn{1}{l}{Previous IDs}\\ 
\multicolumn{1}{l}{} &
\multicolumn{1}{c}{} & \multicolumn{1}{c}{} &
\multicolumn{1}{c}{} & \multicolumn{1}{c}{} &
\multicolumn{1}{c}{} & \multicolumn{2}{c}{(mas yr$^{-1}$)} &
\multicolumn{1}{c}{(\%)} &
\multicolumn{1}{l}{}
\\
\hline
\\
\endfirsthead
\multicolumn{10}{c}%
{{\tablename\ \thetable -- continued from previous page}} \\
\hline \hline \multicolumn{1}{c}{ID} &
\multicolumn{1}{c}{$Z$} & \multicolumn{1}{c}{$Y$} &
\multicolumn{1}{c}{$J$} & \multicolumn{1}{c}{$H$} &
\multicolumn{1}{c}{$K$} & \multicolumn{1}{c}{$\mu_\alpha$} &
\multicolumn{1}{c}{$\mu_\delta$} & \multicolumn{1}{c}{Pmem} &
\multicolumn{1}{l}{Previous IDs}\\ 
\multicolumn{1}{l}{} &
\multicolumn{1}{c}{} & \multicolumn{1}{c}{} &
\multicolumn{1}{c}{} & \multicolumn{1}{c}{} &
\multicolumn{1}{c}{} & \multicolumn{2}{c}{(mas yr$^{-1}$)} &
\multicolumn{1}{c}{(\%)} &
\multicolumn{1}{l}{}
\\
\hline
\\
\endhead
\hline \multicolumn{10}{l}{{Continued on next page}} \\
\endfoot
\hline
\endlastfoot
UGCSJ083545.87+223042. & 14.70 & 14.27 & 13.75 & 13.18 & 12.89 & -35.9 & -11.8 & 0.65 &                    AD2057; KH561 \\ 
UGCSJ084207.83+221105. & 14.84 & 14.39 & 13.84 & 13.27 & 12.99 & -35.5 &  -7.3 & 0.76 &                    AD3054; HSHJ406; KH691 \\ 
UGCSJ083722.41+220200. & 14.11 & 13.61 & 13.05 & 12.52 & 12.21 & -34.0 &  -4.2 & 0.77 &                    AD2305; KH799 \\ 
UGCSJ084440.47+214553. & 14.08 & 13.71 & 13.20 & 12.62 & 12.32 & -34.6 &  -1.3 & 0.71 &                    AD3337; KH508 \\ 
UGCSJ084120.88+215453. & 13.96 & 13.71 & 13.24 & 12.64 & 12.43 & -28.5 & -10.4 & 0.75 &                    AD2939  \\ 
UGCSJ084302.88+214513. & 15.13 & 14.57 & 13.97 & 13.44 & 13.08 & -29.6 &  -9.6 & 0.73 &                    AD3161; KH989  \\ 
UGCSJ083256.66+213829. & 15.82 & 15.43 & 14.92 & 14.36 & 14.05 & -38.2 &  -6.3 & 0.69 &                    AD1687  \\ 
UGCSJ084126.00+213425. & 15.93 & 15.42 & 14.82 & 14.31 & 13.94 & -36.7 &  -3.8 & 0.72 &                    AD2951; HSHJ367; KH901 \\ 
UGCSJ084458.84+213217. & 15.64 & 15.28 & 14.76 & 14.28 & 13.98 & -27.8 &  -0.6 & 0.61 &                    AD3361 \\ 
UGCSJ083454.93+213854. & 15.19 & 14.75 & 14.17 & 13.59 & 13.28 & -34.2 &  -1.4 & 0.71 &                    AD1951; KH773  \\ 
UGCSJ083526.81+213901. & 15.74 & 15.25 & 14.65 & 14.10 & 13.77 & -37.8 &  -8.8 & 0.68 &                    AD2021; KH843 \\ 
UGCSJ083912.55+213557. & 15.90 & 15.43 & 14.81 & 14.24 & 13.93 & -29.2 &  -7.2 & 0.75 &                    AD2517; HSHJ250 \\ 
UGCSJ083413.87+212352. & 15.25 & 14.75 & 14.18 & 13.61 & 13.31 & -27.8 &  -8.1 & 0.71 &                    AD1868; KH786 \\ 
UGCSJ083434.27+212207. & 14.54 & 14.06 & 13.46 & 12.89 & 12.59 & -29.2 &  -5.2 & 0.75 &                    AD1915; KH681 \\ 
UGCSJ083316.62+212020. & 14.08 & 13.64 & 13.11 & 12.54 & 12.26 & -29.8 &  -9.3 & 0.74 &                    AD1737; KH677 \\ 
UGCSJ084143.40+212950. & 15.93 & 15.25 & 14.60 & 14.08 & 13.70 & -31.7 & -12.4 & 0.67 &                    AD3006; HSHJ386; KH1084 \\ 
UGCSJ084030.70+212333. & 14.94 & 14.56 & 14.04 & 13.47 & 13.19 & -29.1 &  -9.0 & 0.73 &                    AD2776; HSHJ318;  \\ 
UGCSJ084048.51+212949. & 13.64 & 13.43 & 13.00 & 12.40 & 12.24 & -26.0 &  -8.8 & 0.81 &                    AD2828 \\ 
UGCSJ082935.64+212047. & 14.79 & 14.53 & 14.07 & 13.45 & 13.26 & -31.2 &  -4.2 & 0.77 &                    AD1252 \\ 
UGCSJ084536.22+211521. & 13.74 & 13.38 & 12.87 & 12.28 & 12.01 & -26.8 &  -1.4 & 0.73 &                    AD3427; HSHJ497; JS604; KH450 \\ 
UGCSJ084457.00+210648. & 15.58 & 15.11 & 14.54 & 13.97 & 13.67 & -28.3 &  -6.2 & 0.73 &                    AD3358; KH835 \\ 
UGCSJ083813.89+210926. & 13.91 & 13.55 & 13.07 & 12.47 & 12.21 & -31.8 &  -9.9 & 0.63 &                    HSHJ198; JS216; KH453  \\ 
UGCSJ083459.25+210837. & 15.42 & 14.84 & 14.24 & 13.69 & 13.37 & -32.9 &  -1.7 & 0.73 &                    AD1962; KH961 \\ 
UGCSJ084123.94+211519. & 15.62 & 15.24 & 14.73 & 14.17 & 13.90 & -28.4 &  -3.9 & 0.72 &                    AD2945 \\ 
UGCSJ084719.06+211102. & 15.58 & 15.11 & 14.51 & 13.95 & 13.64 & -25.2 &  -6.3 & 0.60 &                    KH814 \\ 
UGCSJ084711.92+210748. & 15.72 & 15.22 & 14.60 & 14.07 & 13.74 & -32.4 & -12.3 & 0.67 &                    HSHJ501; KH926 \\ 
UGCSJ083730.73+210740. & 14.94 & 14.51 & 13.99 & 13.42 & 13.13 & -32.7 &  -9.7 & 0.75 &                    KH564 \\ 
UGCSJ082927.94+210838. & 13.39 & 13.11 & 12.65 & 12.04 & 11.80 & -31.1 &  -3.6 & 0.78 &                    AD1240; KH397 \\ 
UGCSJ084515.55+210335. & 14.41 & 14.04 & 13.52 & 12.96 & 12.69 & -36.0 &  -1.9 & 0.70 &                    AD3394; HSHJ496; KH574 \\ 
UGCSJ084620.04+210032. & 15.80 & 15.29 & 14.69 & 14.14 & 13.82 & -36.7 &  -6.4 & 0.73 &                    AD3506; HSHJ499; KH927 \\ 
UGCSJ084321.75+205510. & 15.07 & 14.78 & 14.28 & 13.74 & 13.48 & -26.7 &  -4.8 & 0.67 &                    AD3195 \\ 
UGCSJ082942.64+205707. & 14.03 & 13.78 & 13.28 & 12.70 & 12.48 & -28.4 &  -0.3 & 0.61 &                    AD1263 \\ 
UGCSJ083629.41+210310. & 14.94 & 14.54 & 14.01 & 13.44 & 13.16 & -31.7 & -13.5 & 0.61 &                    AD2175; HSHJ125; KH614 \\ 
UGCSJ083715.24+205759. & 14.20 & 13.90 & 13.42 & 12.82 & 12.59 & -27.5 &  -2.5 & 0.66 &                    AD2291 \\ 
UGCSJ083401.54+210039. & 15.77 & 15.27 & 14.67 & 14.11 & 13.80 & -25.4 &  -5.6 & 0.61 &                    AD1837; KH871 \\ 
UGCSJ084114.43+205946. & 15.63 & 15.17 & 14.58 & 14.02 & 13.71 & -30.7 &  -1.0 & 0.70 &                    AD2918; HSHJ356; KH838 \\ 
UGCSJ084859.88+204155. & 15.69 & 15.12 & 14.50 & 13.99 & 13.62 & -39.3 &  -6.6 & 0.63 &                    KH965 \\ 
UGCSJ083918.03+204421. & 13.67 & 13.36 & 12.86 & 12.26 & 12.05 & -31.6 &  -8.5 & 0.74 &                    AD2538; JS284; KH419 \\ 
UGCSJ083922.13+204758. & 15.27 & 14.78 & 14.20 & 13.65 & 13.32 & -32.3 &  -4.5 & 0.78 &                    AD2551; HSHJ261; KH828 \\ 
UGCSJ083232.42+205040. & 14.45 & 14.06 & 13.53 & 12.95 & 12.70 & -26.5 &  -5.8 & 0.67 &                    AD1632; JS10; KH528 \\ 
UGCSJ083845.67+203943. & 15.44 & 15.02 & 14.43 & 13.90 & 13.60 & -34.5 &  -4.1 & 0.77 &                    AD2452; KH812  \\ 
UGCSJ083615.51+204109. & 13.15 & 12.90 & 12.44 & 11.83 & 11.62 & -31.7 &  -4.8 & 0.78 &                    AD2138; JS117; KH357 \\ 
UGCSJ083603.23+205015. & 15.67 & 15.18 & 14.61 & 14.08 & 13.74 & -33.6 &  -5.7 & 0.78 &                    AD2101; HSHJ096; KH841 \\ 
UGCSJ084114.04+204429. & 13.97 & 13.53 & 12.96 & 12.38 & 12.12 & -26.5 &  -5.2 & 0.87 &                    AD2916; JS416; KH707 \\ 
UGCSJ083711.87+204047. & 14.03 & 13.72 & 13.20 & 12.59 & 12.35 & -27.7 & -11.9 & 0.61 &                    AD2284; JS166l; KH477 \\ 
UGCSJ083406.67+204946. & 15.71 & 15.26 & 14.68 & 14.11 & 13.80 & -29.4 &  -5.6 & 0.76 &                    AD1852; KH818  \\ 
UGCSJ084418.23+204948. & 15.45 & 14.94 & 14.32 & 13.76 & 13.46 & -38.4 &  -1.7 & 0.61 &                    AD3300; HSHJ478 \\ 
UGCSJ083300.38+204310. & 13.72 & 13.35 & 12.79 & 12.23 & 11.95 & -32.0 &  -3.9 & 0.73 &                    AD1699; JS19; KH706 \\ 
UGCSJ083804.60+203935. & 14.95 & 14.53 & 13.95 & 13.38 & 13.08 & -33.0 & -10.4 & 0.74 &                    KH663; JS704; \\ 
UGCSJ084849.96+202635. & 13.44 & 13.15 & 12.66 & 12.07 & 11.81 & -28.7 &  -4.3 & 0.86 &                    KH392 \\ 
UGCSJ085032.50+203419. & 14.27 & 13.97 & 13.51 & 12.99 & 12.81 & -34.0 &  -0.3 & 0.68 &                    AD3783 \\ 
UGCSJ084545.87+202940. & 13.78 & 13.44 & 12.92 & 12.34 & 12.09 & -28.4 &  -8.8 & 0.83 &                    AD3447; JS609; KH460 \\ 
UGCSJ084047.77+202847. & 14.03 & 13.61 & 13.09 & 12.49 & 12.25 & -33.6 &  -7.9 & 0.78 &                    AD2825; KH554 \\ 
UGCSJ083314.23+203621. & 15.71 & 15.20 & 14.57 & 14.03 & 13.72 & -28.5 &  -8.4 & 0.72 &                    AD1731; KH892 \\ 
UGCSJ083808.16+202646. & 13.70 & 13.35 & 12.82 & 12.22 & 12.01 & -30.8 &  -6.5 & 0.83 &                    KH418 \\ 
UGCSJ084611.69+203800. & 14.36 & 14.04 & 13.50 & 12.99 & 12.71 & -32.2 &  -6.8 & 0.79 &                    AD3490 \\ 
UGCSJ083338.36+202852. & 15.31 & 14.82 & 14.25 & 13.71 & 13.43 & -27.1 & -10.7 & 0.62 &                    AD1786; KH724  \\ 
UGCSJ083943.59+202939. & 14.54 & 14.14 & 13.60 & 13.03 & 12.75 & -25.7 &  -5.7 & 0.63 &                    AD2618; JS311; KH629 \\ 
UGCSJ083912.08+203607. & 14.84 & 14.38 & 13.82 & 13.27 & 12.96 & -39.3 &  -6.0 & 0.64 &                    AD2515 \\ 
UGCSJ083903.93+203402. & 13.98 & 13.60 & 13.07 & 12.50 & 12.24 & -25.8 &  -7.4 & 0.84 &                    AD2502; JS266; KH500 \\ 
UGCSJ082750.59+201436. & 12.50 & 12.26 & 11.85 & 11.32 & 11.09 & -26.0 &  -0.4 & 0.60 &                    AD1025; KH300 \\ 
UGCSJ084111.05+202238. & 13.44 & 13.12 & 12.63 & 12.04 & 11.81 & -25.7 &  -5.8 & 0.85 &                    AD2905; JS411; KH416 \\ 
UGCSJ084137.35+201236. & 15.15 & 14.66 & 14.09 & 13.56 & 13.27 & -30.2 &  -1.2 & 0.69 &                    AD2988; HSHJ381; KH667 \\ 
UGCSJ083641.16+201639. & 15.04 & 14.60 & 13.98 & 13.45 & 13.14 & -30.7 &  -5.6 & 0.78 &                    AD2205; KH693  \\ 
UGCSJ083642.16+201622. & 15.46 & 14.97 & 14.37 & 13.85 & 13.52 & -32.3 &  -9.9 & 0.75 &                    AD2208 \\ 
UGCSJ084423.19+201355. & 15.34 & 14.84 & 14.28 & 13.72 & 13.42 & -29.1 &  -4.8 & 0.75 &                    AD3312; KH787 \\ 
UGCSJ083041.51+202426. & 15.55 & 15.17 & 14.68 & 14.10 & 13.84 & -37.2 &  -9.0 & 0.69 &                    AD1396 \\ 
UGCSJ083311.09+201604. & 15.16 & 14.74 & 14.17 & 13.67 & 13.37 & -31.8 & -12.6 & 0.66 &                    AD1719 \\ 
UGCSJ083942.01+201745. & 14.98 & 14.50 & 13.93 & 13.39 & 13.11 & -31.5 &  -0.8 & 0.70 &                    AD2615; KH808 \\ 
UGCSJ083906.87+202054. & 13.54 & 13.23 & 12.75 & 12.14 & 11.94 & -25.1 &  -5.5 & 0.83 &                    AD2508; JS270; KH426 \\ 
UGCSJ083507.87+202023. & 14.46 & 14.07 & 13.55 & 12.99 & 12.71 & -30.9 &  -9.5 & 0.75 &                    AD1978; KH496 \\ 
UGCSJ083436.75+201155. & 13.71 & 13.41 & 12.93 & 12.32 & 12.08 & -30.4 &  -6.4 & 0.84 &                    AD1921; HSHJ058; JS63 \\ 
UGCSJ084151.90+202047. & 13.85 & 13.51 & 13.00 & 12.41 & 12.18 & -29.9 &  -8.4 & 0.82 &                    AD3026; JS459; KH470 \\ 
UGCSJ083855.15+201308. & 14.31 & 13.90 & 13.35 & 12.77 & 12.50 & -32.3 &   0.1 & 0.67 &                    AD2478; HSHJ234; JS255; KH606 \\ 
UGCSJ083825.35+202120. & 15.93 & 15.45 & 14.85 & 14.32 & 13.99 & -31.1 & -11.8 & 0.69 &                    KH903 \\ 
UGCSJ083539.26+202409. & 14.98 & 14.48 & 13.88 & 13.35 & 13.04 & -31.4 &  -8.2 & 0.77 &                    AD2042; KH854 \\ 
UGCSJ085237.29+200043. & 13.46 & 13.26 & 12.84 & 12.27 & 12.13 & -26.4 &  -5.0 & 0.86 &                    AD4003 \\ 
UGCSJ083259.56+200714. & 15.87 & 15.39 & 14.81 & 14.27 & 13.93 & -33.9 & -12.6 & 0.65 &                    AD1696; KH896 \\ 
UGCSJ084332.60+195932. & 14.73 & 14.31 & 13.75 & 13.19 & 12.88 & -34.2 &  -3.9 & 0.77 &                    AD3211; HSHJ458; KH655 \\ 
UGCSJ083333.93+200425. & 14.27 & 13.89 & 13.36 & 12.76 & 12.52 & -26.6 &  -6.4 & 0.67 &                    AD1774; HSHJ043; KH509  \\ 
UGCSJ083559.43+200440. & 14.87 & 14.41 & 13.81 & 13.26 & 12.95 & -36.5 &  -1.0 & 0.66 &                    AD2091; HSHJ091; JS687; KH743 \\ 
UGCSJ083622.40+200706. & 15.36 & 14.89 & 14.32 & 13.76 & 13.45 & -26.3 &  -7.9 & 0.65 &                    AD2155; KH774  \\ 
UGCSJ084034.83+194937. & 14.85 & 14.52 & 14.04 & 13.46 & 13.19 & -34.7 &  -5.9 & 0.77 &                    AD2790 \\ 
UGCSJ083608.58+195725. & 15.13 & 14.65 & 14.05 & 13.49 & 13.19 & -26.1 &  -9.0 & 0.62 &                    AD2115; HSHJ102; KH884 \\ 
UGCSJ083619.14+195354. & 14.34 & 13.94 & 13.41 & 12.82 & 12.56 & -31.4 &  -8.2 & 0.77 &                    AD2147; HSHJ115; JS123; KH593 \\ 
UGCSJ083707.63+195727. & 15.47 & 14.99 & 14.43 & 13.87 & 13.56 & -32.5 &  -7.8 & 0.78 &                    AD2275; KH864 \\ 
UGCSJ083727.87+195412. & 14.32 & 13.91 & 13.38 & 12.78 & 12.53 & -32.7 & -11.3 & 0.71 &                    AD2328; KH638 \\ 
UGCSJ084030.57+195558. & 14.15 & 13.76 & 13.25 & 12.64 & 12.42 & -27.0 & -10.5 & 0.63 &                    AD2775; HSHJ320; JS356; KH546  \\ 
UGCSJ084801.27+194939. & 15.40 & 14.95 & 14.37 & 13.79 & 13.50 & -32.5 &  -5.6 & 0.79 &                    KH727 \\ 
UGCSJ084823.56+195011. & 14.64 & 14.23 & 13.65 & 13.07 & 12.78 & -31.9 &   1.1 & 0.62 &                    KH579 \\ 
UGCSJ085056.86+193657. & 14.44 & 14.05 & 13.52 & 12.93 & 12.67 & -30.7 &   0.6 & 0.63 &                    KH527 \\ 
UGCSJ082757.39+191130. & 15.94 & 15.55 & 14.98 & 14.37 & 14.12 & -36.2 &  -9.1 & 0.72 &                    AD1043 \\ 
UGCSJ083439.68+190812. & 14.61 & 14.18 & 13.63 & 13.05 & 12.78 & -32.7 &  -4.5 & 0.78 &                    AD1925; HSHJ060; KH596 \\ 
UGCSJ083430.71+190600. & 14.46 & 14.08 & 13.53 & 12.96 & 12.73 & -38.2 &  -9.9 & 0.63 &                    AD1906; HSHJ055; JS675; KH513 \\ 
UGCSJ082848.63+185835. & 15.82 & 15.26 & 14.63 & 14.07 & 13.72 & -31.5 &  -5.9 & 0.78 &                    AD1164; KH1025 \\ 
UGCSJ083150.86+185902. & 15.29 & 14.85 & 14.32 & 13.79 & 13.47 & -29.9 &   0.6 & 0.62 &                    AD1549 \\ 
UGCSJ083218.87+190308. & 14.64 & 14.21 & 13.64 & 13.08 & 12.78 & -33.6 &  -8.4 & 0.77 &                    AD1607; HSHJ023; KH594  \\ 
UGCSJ083544.59+185738. & 15.40 & 14.93 & 14.35 & 13.80 & 13.50 & -29.2 &  -6.0 & 0.75 &                    AD2053; HSHJ084; KH858 \\ 
UGCSJ083651.05+190418. & 14.99 & 14.61 & 14.09 & 13.51 & 13.22 & -28.7 &  -4.4 & 0.73 &                    AD2231; HSHJ136; KH565 \\ 
UGCSJ083305.56+185548. & 13.93 & 13.55 & 13.01 & 12.43 & 12.17 & -22.8 &  -4.3 & 0.66 &                    AD1707; JS22; KH519 \\ 
UGCSJ083338.00+185717. & 14.21 & 13.86 & 13.31 & 12.74 & 12.49 & -38.5 &  -7.4 & 0.67 &                    AD1785; HSHJ045; JS41; KH482 \\ 
UGCSJ083051.38+185351. & 15.25 & 14.81 & 14.22 & 13.66 & 13.38 & -32.3 &   0.0 & 0.67 &                    AD1411; HSHJ008; KH670 \\ 
UGCSJ083734.95+185607. & 14.47 & 14.11 & 13.60 & 13.09 & 12.84 & -32.2 &  -6.9 & 0.79 &                    HSHJ173 \\ 
UGCSJ083235.23+184409. & 15.94 & 15.41 & 14.79 & 14.24 & 13.95 & -27.6 &  -6.1 & 0.71 &                    AD1641; HSHJ024 \\ 
UGCSJ083207.94+184426. & 14.37 & 13.97 & 13.47 & 12.87 & 12.59 & -26.9 &  -1.6 & 0.61 &                    AD1588; KH511 \\ 
UGCSJ083154.25+184536. & 15.59 & 15.08 & 14.51 & 13.96 & 13.64 & -28.9 &  -5.9 & 0.75 &                    AD1555; HSHJ018; KH834 \\ 
UGCSJ083126.88+184056. & 15.41 & 14.84 & 14.19 & 13.64 & 13.30 & -29.9 &  -6.8 & 0.77 &                    AD1500; KH1044  \\ 
UGCSJ083658.63+184952. & 15.19 & 14.63 & 14.04 & 13.51 & 13.17 & -28.2 &  -7.6 & 0.72 &                    AD2248; KH923 \\ 
UGCSJ083808.07+184429. & 15.80 & 15.28 & 14.68 & 14.11 & 13.82 & -27.0 & -10.5 & 0.63 &                    HSHJ196; KH779 \\ 
UGCSJ083729.39+184135. & 14.89 & 14.48 & 13.94 & 13.38 & 13.10 & -37.7 &  -7.2 & 0.70 &                    HSHJ165; KH542 \\ 
UGCSJ083528.37+184032. & 15.73 & 15.21 & 14.63 & 14.10 & 13.76 & -27.3 &  -3.6 & 0.68 &                    AD2022 \\ 
UGCSJ083506.21+184924. & 14.76 & 14.33 & 13.77 & 13.20 & 12.92 & -38.7 &  -3.8 & 0.64 &                    AD1975; HSHJ068; JS680; KH644 \\ 
UGCSJ083540.13+184228. & 14.15 & 13.72 & 13.18 & 12.62 & 12.33 & -34.2 &  -5.9 & 0.78 &                    AD2045; HSHJ082; JS95; KH636  \\ 
UGCSJ083434.30+184756. & 14.13 & 13.73 & 13.19 & 12.62 & 12.37 & -32.5 & -12.6 & 0.66 &                    AD1917; HSHJ056; JS62; KH570 \\ 
UGCSJ083343.93+184750. & 14.31 & 13.90 & 13.40 & 12.81 & 12.54 & -31.9 & -13.7 & 0.60 &                    AD1797; HSHJ047; JS44; KH526  \\ 
UGCSJ083328.19+184336. & 15.96 & 15.40 & 14.81 & 14.28 & 13.95 & -38.4 &  -4.3 & 0.67 &                    AD1763; HSHJ041; KH874 \\ 
UGCSJ083142.95+182906. & 15.26 & 14.81 & 14.22 & 13.64 & 13.36 & -26.2 &  -7.9 & 0.65 &                    AD1532; HSHJ017; KH699  \\ 
UGCSJ083140.88+182942. & 14.89 & 14.47 & 13.90 & 13.34 & 13.08 & -32.9 &  -5.7 & 0.79 &                    AD1526; HSHJ016; KH563 \\ 
UGCSJ083521.67+182934. & 13.98 & 13.59 & 13.11 & 12.53 & 12.28 & -27.0 &  -7.6 & 0.86 &                    AD2011; HSHJ076; JS87; KH463 \\ 
UGCSJ083334.78+183108. & 13.24 & 12.97 & 12.54 & 11.98 & 11.84 & -26.6 &  -3.3 & 0.83 &                    AD1776; JS40 \\ 
UGCSJ083014.08+182519. & 14.56 & 14.15 & 13.61 & 13.08 & 12.78 & -35.1 & -11.9 & 0.66 &                    AD1344; HSHJ004; KH559 \\ 
UGCSJ083453.83+180105. & 15.72 & 15.23 & 14.61 & 14.05 & 13.73 & -32.0 & -10.9 & 0.72 &                    AD1948; HSHJ065; KH817 \\ 
UGCSJ083547.20+180829. & 14.54 & 14.12 & 13.52 & 12.94 & 12.67 & -30.7 & -11.7 & 0.69 &                    AD2063; HSHJ087; KH628  \\ 
UGCSJ083002.92+175702. & 15.79 & 15.24 & 14.65 & 14.07 & 13.72 & -32.0 & -12.0 & 0.69 &                    AD1308; HSHJ002; KH990 \\ 
UGCSJ083517.03+173624. & 15.12 & 14.59 & 14.00 & 13.43 & 13.14 & -27.1 &  -7.7 & 0.68 &                    AD2000; HSHJ074; KH855 \\ 
UGCSJ083839.13+172948. & 15.98 & 15.49 & 14.88 & 14.33 & 14.02 & -30.6 & -10.7 & 0.72 &                    AD2436; HSHJ217; KH904 \\ 
UGCSJ084026.64+172100. & 14.82 & 14.48 & 13.98 & 13.40 & 13.13 & -29.9 &  -2.9 & 0.73 &                    AD2760 \\ 
UGCSJ083855.64+171509. & 14.51 & 14.12 & 13.56 & 12.99 & 12.70 & -27.4 &  -2.3 & 0.65 &                    AD2482; KH608 \\ 
UGCSJ083824.87+165836. & 14.09 & 13.71 & 13.16 & 12.59 & 12.31 & -32.8 &  -2.0 & 0.74 &                    AD2396; KH478 \\ 
UGCSJ083906.50+170100. & 14.58 & 14.06 & 13.51 & 12.96 & 12.64 & -36.2 &  -7.5 & 0.74 &                    AD2507 \\ 
UGCSJ083608.51+165717. & 13.88 & 13.54 & 13.05 & 12.50 & 12.22 & -29.8 &  -4.0 & 0.84 &                    AD2114 \\ 

\\
\end{longtable}
\end{center}

\twocolumn
\normalsize

\onecolumn
\section{New High Probability Members}
\label{HPM_new_Appendix}
\scriptsize
\begin{center}
\begin{longtable}{ccccccccc}
\caption{New High Probability Members}\\
\hline \hline \multicolumn{1}{c}{ID} &
\multicolumn{1}{c}{$Z$} & \multicolumn{1}{c}{$Y$} &
\multicolumn{1}{c}{$J$} & \multicolumn{1}{c}{$H$} &
\multicolumn{1}{c}{$K$} & \multicolumn{1}{c}{$\mu_\alpha$} &
\multicolumn{1}{c}{$\mu_\delta$} & \multicolumn{1}{c}{Pmem}\\ 
\multicolumn{1}{l}{} &
\multicolumn{1}{c}{} & \multicolumn{1}{c}{} &
\multicolumn{1}{c}{} & \multicolumn{1}{c}{} &
\multicolumn{1}{c}{} & \multicolumn{2}{c}{(mas yr$^{-1}$)} &
\multicolumn{1}{c}{}
\\
\hline
\\
\endfirsthead
\multicolumn{9}{c}%
{{\tablename\ \thetable -- continued from previous page}} \\
\hline \hline \multicolumn{1}{c}{ID} &
\multicolumn{1}{c}{$Z$} & \multicolumn{1}{c}{$Y$} &
\multicolumn{1}{c}{$J$} & \multicolumn{1}{c}{$H$} &
\multicolumn{1}{c}{$K$} & \multicolumn{1}{c}{$\mu_\alpha$} &
\multicolumn{1}{c}{$\mu_\delta$} & \multicolumn{1}{c}{Pmem}\\ 
\multicolumn{1}{l}{} &
\multicolumn{1}{c}{} & \multicolumn{1}{c}{} &
\multicolumn{1}{c}{} & \multicolumn{1}{c}{} &
\multicolumn{1}{c}{} & \multicolumn{2}{c}{(mas yr$^{-1}$)} &
\multicolumn{1}{c}{}\\
\hline
\\
\endhead
\hline \multicolumn{9}{l}{{Continued on next page}} \\
\endfoot
\hline
\endlastfoot
 UGCSJ083925.46+214721.9 & 15.46 & 15.12 & 14.62 & 14.06 & 13.81 & -31.1 &  -8.8 & 0.76 \\
 UGCSJ083301.71+213902.1 & 13.95 & 13.75 & 13.35 & 12.73 & 12.58 & -24.6 &  -4.5 & 0.80 \\
 UGCSJ084826.25+213235.7 & 15.98 & 15.59 & 15.02 & 14.48 & 14.20 & -32.4 &  -1.6 & 0.73 \\
 UGCSJ083310.88+210959.8 & 13.38 & 13.15 & 12.71 & 12.10 & 11.92 & -22.9 &  -6.5 & 0.70 \\
 UGCSJ083032.17+211015.4 & 15.26 & 14.96 & 14.46 & 13.84 & 13.66 & -29.4 &  -7.3 & 0.76 \\
 UGCSJ083956.01+211419.1 & 15.84 & 15.49 & 15.00 & 14.43 & 14.18 & -25.5 &  -4.8 & 0.61 \\
 UGCSJ083553.75+191055.8 & 15.11 & 14.85 & 14.39 & 13.79 & 13.61 & -26.2 &  -6.3 & 0.65 \\
 UGCSJ083115.62+184708.9 & 15.89 & 15.58 & 15.13 & 14.57 & 14.35 & -31.1 &  -0.3 & 0.68 \\
 UGCSJ083448.26+181300.2 & 14.46 & 14.26 & 13.82 & 13.21 & 13.08 & -27.6 &  -4.5 & 0.70 \\
 UGCSJ083459.25+181805.4 & 15.43 & 15.19 & 14.71 & 14.10 & 13.88 & -32.9 &  -1.5 & 0.72 \\
 UGCSJ083306.96+174242.1 & 15.29 & 14.77 & 14.16 & 13.61 & 13.31 & -26.3 &  -3.6 & 0.63 \\
 UGCSJ083701.06+172005.1 & 14.80 & 14.47 & 13.94 & 13.41 & 13.12 & -28.1 &  -7.2 & 0.72 \\
 UGCSJ084709.34+172925.6 & 15.45 & 15.11 & 14.58 & 14.12 & 13.87 & -33.1 &  -9.3 & 0.76 \\
 UGCSJ083455.71+170918.5 & 13.90 & 13.72 & 13.30 & 12.70 & 12.56 & -22.6 &  -4.1 & 0.63 \\

\\
\end{longtable}
\end{center}

\twocolumn
\normalsize

\onecolumn
\section{SDSS Candidate Members}
 \footnote{For some objects not enough reference stars could be found to perform the quadratic fit. In this case the fit was reduced to a linear one (6 free parameters) as indicated in the fit order column.}
\label{SDSS_cands_appendix}
\scriptsize
\begin{center}
\begin{longtable}{cccccccccc}
\caption{SDSS-UKIDSS selected candidates}\\
\hline \hline \multicolumn{1}{c}{ID} &
\multicolumn{1}{c}{$Z$} & \multicolumn{1}{c}{$Y$} &
\multicolumn{1}{c}{$J$} & \multicolumn{1}{c}{$H$} &
\multicolumn{1}{c}{$K$} & \multicolumn{1}{c}{$\mu_\alpha$} &
\multicolumn{1}{c}{$\mu_\delta$} & \multicolumn{1}{c}{Prob} &
\multicolumn{1}{c}{Fit Order}\\ 
\multicolumn{1}{c}{} &
\multicolumn{1}{c}{} & \multicolumn{1}{c}{} &
\multicolumn{1}{c}{} & \multicolumn{1}{c}{} &
\multicolumn{1}{c}{} & \multicolumn{2}{c}{(mas yr$^{-1}$)} &
\multicolumn{1}{c}{} & \multicolumn{1}{c}{} \\
\hline
\\
\endfirsthead
\multicolumn{10}{c}%
{{\tablename\ \thetable -- continued from previous page}} \\
\hline \hline \multicolumn{1}{c}{ID} &
\multicolumn{1}{c}{$Z$} & \multicolumn{1}{c}{$Y$} &
\multicolumn{1}{c}{$J$} & \multicolumn{1}{c}{$H$} &
\multicolumn{1}{c}{$K$} & \multicolumn{1}{c}{$\mu_\alpha$} &
\multicolumn{1}{c}{$\mu_\delta$} & \multicolumn{1}{c}{Prob} &
\multicolumn{1}{c}{Fit Order}\\ 
\multicolumn{1}{c}{} &
\multicolumn{1}{c}{} & \multicolumn{1}{c}{} &
\multicolumn{1}{c}{} & \multicolumn{1}{c}{} &
\multicolumn{1}{c}{} & \multicolumn{2}{c}{(mas yr$^{-1}$)} &
\multicolumn{1}{c}{} & \multicolumn{1}{c}{} \\
\hline
\\
\endhead
\hline \multicolumn{10}{l}{{Continued on next page}} \\
\endfoot
\hline
\endlastfoot
 UGCSJ084218.27+212342.1 & 18.18 & 17.50 & 16.81 & 16.27 & 15.88 &    -4.05 &    -6.32 & 0.00 & 12 \\
 UGCSJ084804.97+211515.1 & 19.64 & 18.57 & 17.68 & 17.05 & 16.59 &    67.66 &    64.70 & 0.00 &  6 \\
 UGCSJ084346.46+210829.4 & 18.46 & 17.68 & 17.00 & 16.45 & 16.06 &   -28.38 &   -38.64 & 0.00 & 12 \\
 UGCSJ084956.26+205300.0 & 17.88 & 17.30 & 16.65 & 16.09 & 15.71 &    32.59 &   -53.01 & 0.00 & 12 \\
 UGCSJ082924.86+211700.3 & 16.10 & 15.59 & 15.02 & 14.57 & 14.27 &   -94.95 &  -133.65 & 0.00 & 12 \\
 UGCSJ084138.84+211655.5 & 17.88 & 17.29 & 16.55 & 16.03 & 15.72 &     9.02 &    20.03 & 0.00 & 12 \\
 UGCSJ084449.14+210153.6 & 18.77 & 17.79 & 16.95 & 16.41 & 15.94 &     0.36 &    23.91 & 0.00 & 12 \\
 UGCSJ084507.41+210056.5 & 18.69 & 17.87 & 17.06 & 16.49 & 16.14 &   -36.74 &   -40.16 & 0.00 & 12 \\
 UGCSJ084204.64+205941.1 & 16.77 & 16.09 & 15.42 & 14.93 & 14.54 &   114.67 &  -167.13 & 0.00 & 12 \\
 UGCSJ082839.58+205401.5 & 18.48 & 17.69 & 16.91 & 16.34 & 15.95 &    15.24 &   -31.20 & 0.00 & 12 \\
 UGCSJ082945.32+205455.0 & 19.26 & 18.48 & 17.67 & 17.10 & 16.51 &     4.16 &   -22.41 & 0.00 & 12 \\
 UGCSJ083054.47+190119.4 & 16.13 & 15.47 & 14.76 & 14.21 & 13.84 &   -15.32 &  -128.89 & 0.00 & 12 \\
 UGCSJ084433.77+214430.7 & 17.77 & 17.01 & 16.30 & 15.74 & 15.32 &   -37.33 &    32.77 & 0.00 &  6 \\
 UGCSJ084650.44+214805.4 & 17.37 & 16.68 & 16.08 & 15.62 & 15.28 &   -75.34 &  -117.60 & 0.00 & 12 \\
 UGCSJ083301.71+213534.7 & 18.65 & 17.93 & 17.20 & 16.61 & 16.17 &    14.87 &    18.57 & 0.00 &  6 \\
 UGCSJ084306.24+214134.2 & 18.40 & 17.56 & 16.71 & 16.16 & 15.73 &   -38.63 &   -44.50 & 0.00 & 12 \\
 UGCSJ083141.80+183500.4 & 15.65 & 15.07 & 14.45 & 13.98 & 13.64 &    46.31 &  -110.20 & 0.00 & 12 \\
 UGCSJ083353.37+182609.4 & 19.20 & 18.17 & 17.24 & 16.64 & 16.18 &     6.73 &    -8.95 & 0.00 & 12 \\
 UGCSJ083955.10+222300.8 & 19.06 & 18.11 & 17.27 & 16.64 & 16.21 &   -11.29 &    -9.27 & 0.58 & 12 \\
 UGCSJ083110.12+181252.7 & 18.15 & 17.48 & 16.74 & 16.20 & 15.75 &    38.81 &   -45.95 & 0.00 & 12 \\
 UGCSJ083654.60+195415.7 & 18.75 & 17.96 & 17.10 & 16.53 & 15.96 &   -26.97 &    28.33 & 0.00 & 12 \\
 UGCSJ083648.03+194902.2 & 18.21 & 17.34 & 16.55 & 15.97 & 15.52 &   -98.98 &   -17.63 & 0.00 &  6 \\
 UGCSJ084714.47+194643.8 & 18.06 & 17.45 & 16.80 & 16.27 & 15.86 &    14.68 &     6.35 & 0.00 & 12 \\
 UGCSJ085052.56+195321.9 & 19.24 & 18.34 & 17.40 & 16.75 & 16.22 &    -5.74 &   -21.30 & 0.00 & 12 \\
 UGCSJ084652.65+172017.9 & 18.67 & 17.91 & 17.15 & 16.62 & 16.18 &    16.12 &   -30.63 & 0.00 & 12 \\
 UGCSJ084045.71+171218.3 & 19.09 & 18.17 & 17.20 & 16.60 & 16.17 &   -33.60 &     1.04 & 0.58 & 12 \\
 UGCSJ084452.78+171409.8 & 18.77 & 17.76 & 16.90 & 16.33 & 15.84 &   -44.67 &   -13.78 & 0.00 & 12 \\
 UGCSJ083900.61+204356.2 & 20.05 & 18.76 & 17.91 & 17.33 & 16.72 &     2.62 &   -71.34 & 0.00 & 12 \\
 UGCSJ083920.28+204629.5 & 19.34 & 18.49 & 17.65 & 16.95 & 16.51 &    -5.94 &   -12.50 & 0.00 & 12 \\
 UGCSJ083721.24+204306.1 & 19.73 & 18.66 & 17.56 & 16.84 & 16.33 &    25.55 &     9.35 & 0.00 & 12 \\
 UGCSJ083737.24+202902.6 & 19.71 & 18.60 & 17.39 & 16.64 & 16.04 &   -34.02 &   -61.35 & 0.00 & 12 \\
 UGCSJ083019.80+203418.8 & 18.95 & 17.92 & 17.11 & 16.59 & 16.13 &   -10.23 &   -93.17 & 0.00 & 12 \\
 UGCSJ083724.09+164812.9 & 18.55 & 17.81 & 17.10 & 16.53 & 16.10 &    -5.67 &    11.58 & 0.00 & 12 \\
 UGCSJ083954.96+202955.7 & 18.90 & 17.91 & 17.06 & 16.44 & 15.98 &    33.38 &   -54.56 & 0.00 & 12 \\
 UGCSJ084926.31+202127.0 & 18.64 & 17.76 & 17.00 & 16.54 & 16.08 &   -75.67 &   -30.17 & 0.00 & 12 \\
 UGCSJ085002.02+201725.9 & 19.28 & 18.56 & 17.64 & 17.05 & 16.53 &   -44.04 &    50.96 & 0.00 & 12 \\
 UGCSJ083203.66+202035.8 & 19.90 & 18.68 & 17.74 & 17.04 & 16.52 &   -31.90 &   -59.39 & 0.00 & 12 \\
 UGCSJ083531.32+201838.9 & 18.39 & 17.62 & 16.96 & 16.39 & 15.99 &   -17.01 &    20.43 & 0.00 & 12 \\
 UGCSJ083748.01+201448.5 & 18.09 & 17.25 & 16.51 & 15.94 & 15.53 &   -36.17 &   -44.57 & 0.00 & 12 \\

\\
\end{longtable}
\end{center}
\label{lastpage}
\bsp
\end{document}